\documentclass[twocolumn]{aastex62}

\graphicspath{{./}{figures/}}

\submitjournal{ApJ}

\shorttitle{Blue stragglers and dynamical ages of LMC clusters}
\shortauthors{C. Li et al.}

\begin{document}

\title{Blue straggler stars beyond the Milky Way. IV. Radial 
distributions and dynamical implications}

\correspondingauthor{Chengyuan Li}
\email{chengyuan.li@mq.edu.au}

\author{Chengyuan Li} 
\affil{Department of Physics and Astronomy, Macquarie University,
  Balaclava Road, Sydney, NSW 2109, Australia}
\affil{Research Centre for Astronomy, Astrophysics and Astrophotonics,
  Macquarie University, Balaclava Road, Sydney, NSW 2109, Australia}
\affil{Department of Astronomy, China West Normal University, Nanchong
  637002, China}

\author{Weijia Sun}
\affiliation{Department of Astronomy, Peking University, Yi He Yuan Lu
  5, Hai Dian District, Beijing 100871, China}
\affiliation{Kavli Institute for Astronomy and Astrophysics, Peking
  University, Yi He Yuan Lu 5, Hai Dian District, Beijing 100871,
  China}

\author{Jongsuk Hong}
\affiliation{Kavli Institute for Astronomy and Astrophysics, Peking
  University, Yi He Yuan Lu 5, Hai Dian District, Beijing 100871,
  China}

\author{Licai Deng}
\affiliation{Key Laboratory for Optical Astronomy, National
  Astronomical Observatories, Chinese Academy of Sciences, 20A Datun
  Road, Beijing 100012, China}

\author{Richard de Grijs}
\affiliation{Department of Physics and Astronomy, Macquarie
  University, Balaclava Road, Sydney, NSW 2109, Australia}
\affiliation{Research Centre for Astronomy, Astrophysics and
  Astrophotonics, Macquarie University, Balaclava Road, Sydney, NSW
  2109, Australia}
\affiliation{International Space Science Institute--Beijing, 1
  Nanertiao, Zhongguancun, Hai Dian District, Beijing 100190, China}

\author{Alison Sills}
\affiliation{Department of Physics and Astronomy, McMaster University,
  1280 Main Street West, Hamilton, ON, L8S 4M1 Canada}

\begin{abstract}
Blue straggler stars have been proposeed as powerful indicators to
measure the dynamical state of Galactic globular clusters. Here we
examine for the first time if this framework of blue straggler stars
as dynamical clocks, which was specifically developed for Galactic
globular clusters, may also hold for younger globular clusters in the
Large Magellanic Cloud. Using {\sl Hubble Space Telescope}
observations, we study seven Large Magellanic Cloud star clusters with
ages between $\sim$700 Myr and 7 Gyr. We confirm that our sample
clusters are consistent with the empirical correlation previously
derived for Galactic globular clusters, within a modest tolerance. We
suggest that to further examine if blue straggler stars can measure
the dynamical state of their host clusters over an extended range of
dynamical ages, more studies of dynamically older Magellanic Cloud
clusters are required. We discuss the physical implications of our
results in terms of their central, dimensionless King potential, as
well as the initial retention fraction of black holes.
\end{abstract}

\keywords{blue stragglers --- galaxies: star clusters --- Magellanic
  Clouds --- stars: kinematics and dynamics}

\section{Introduction} \label{S1}

In dense stellar systems like globular clusters (GCs), blue straggler
stars (BSSs) represent examples of the bluest and brightest
populations. They lie on an extension of the main-sequence (MS)
\citep[e.g.,][]{Sand53a,Ferr93a,Li13b} in the color--magnitude diagram
(CMD). BSSs are thought to be more massive than the bulk of the
`normal' stars \citep[e.g.,][]{Fior14a} in star clusters. They are
exotic objects which may have been produced through active stellar
dynamics --- either through mergers of binary components
\citep{Hill76a,Andr06a} or through direct collisions in dense clusters
\citep{Mccr64a}. Therefore, BSSs can in principle reveal information
about the dynamical processes in star clusters.

BSSs were found to be ideal probes of the dynamical history of
Galactic GCs \citep[e.g.,][]{Ales16a,Lanz16a,Ferr18a}. The detection
of bifurcated BSS populations in some post-core-collapse clusters
strongly indicates that stellar collisions driven by cluster core
collapse could produce BSSs within a short period of time
\citep[e.g.,][]{Ferr09a,Dale13a,Simu14a}. \cite{Ferr12a,Lanz16a}
derived an empirical correlation between the radial distributions of
BSSs and the dynamical ages of their host clusters in the Milky Way,
which was recently underpinned by \cite{Ferr18a}. These authors
concluded that the radial distributions of BSSs can serve as a
`dynamical clock' for measuring the dynamical states reached by GCs.

Since BSSs are closely related to binary evolutions, their spatial
segregation may show consistency with that of a cluster's binary
population. An exploration of this type of correlation has been
carried out over the last decade: the radial distribution of binary
systems is indeed similar to that of the BSSs in star clusters
\citep{Milo12a,Gell13a}. In some young massive clusters, the radial
behavior of the binary systems is more complicated, because the
ongoing dynamical binary disruption may mask their mass segregation
\citep{Li13a,Gell15a,Yang18a}. It is thus interesting to explore the
radial behavior of BSSs in clusters younger than most Galactic GCs
($\sim$10 Gyr).

Although the framework describing how BSSs can probe the dynamical
processes of their host stellar systems has been well-studied for
Galactic GCs, whether our understanding of BSSs would also hold for
their younger, extragalactic counterparts remains unclear. A limited
number of studies aiming to explore BSSs in star clusters in the
Magellanic Clouds have recently been carried out. \cite{Li13b}
detected two separated groups of BSSs in the CMD of the Large
Magellanic Cloud (LMC) GC Hodge 11, which is similar (but less
significant) to the bifurcations found in some Galactic GCs
\citep{Ferr09a,Dale13a,Simu14a}. Recently, the young (1--2 Gyr-old)
LMC cluster NGC 2173 was found to exhibit two well-separated BSS
sequences in its CMD \citep{Li18b}, a feature which is commonly
explained as the result of a cluster's core collapse. However, the
number density profile of NGC 2173 does not exhibit a central cusp, a
typical feature created by a collapsed core. For another 1--2 Gyr-old
cluster, NGC 2213, \cite{Li18a} found that the radial distribution of
its BSSs is consistent with that of the normal stars, showing no
evidence of dynamical mass segregation (although dynamical
calculations carried out for that cluster showed that the radial
distribution of its BSSs should have been shaped by dynamical mass
segregation). \cite{Li18a} suggested that this may be because the
presence of some extremely compact objects such as black holes in the
cluster's central region could have delayed the dynamical evolution of
the BSSs. \cite{Sun18a} studied BSSs in 25 Magellanic Cloud star
clusters. They revealed a sublinear correlation between the number of
BSSs in the cluster cores and the clusters' core masses, i.e., $N_{\rm
  BSS,c}\propto{M_{\rm c}}^{0.66\pm0.07}$, where $N_{\rm BSS,c}$ and
$M_{\rm c}$ are the number of BSSs in the core region and the
cluster's core mass, respectively. This feature had been previously
confirmed in Galactic GCs \citep{Knig09a}, which implied that binary
evolution is the major formation channel of BSSs.

In this paper, we study seven LMC massive clusters with ages from
$\leq$1 Gyr to $\sim$7 Gyr. Using diagnostic diagrams similar to those
used by \cite{Lanz16a}, we examine if the BSS mass segregation degrees
could be applied to measure the dynamical states of their host
clusters. The effects of (small-)number dispersions of the BSSs as
well as of field contamination are carefully explored by employing a
Monte-Carlo based statistical examination. We will show that our
result is generally consistent with the empirical relationship derived
by \cite{Lanz16a} and \cite{Ferr18a} for Galactic GCs. We discuss the
physical implications of our results by comparing our observations
with the numerical simulation results of \cite{Ales16a}.

We introduce the details of the observations and the data reduction in
Section \ref{S2}. The main results of our analyses are presented in
Section \ref{S3}. In Section \ref{S4}, we provide a discussion of the
physical implications of our results. Section \ref{S5} contains a
summary and our conclusions.

\section{Data Reduction} \label{S2}
\subsection{Star cluster selection} \label{S2a}

We first examined all LMC clusters contained in the {\sl Hubble Space
  Telescope} ({\sl HST}) archive data
set\footnote{\url{https://archive.stsci.edu/hst/search.php}} using the
cluster catalog of \cite{Baum13a}. We selected data sets observed with
the {\sl HST} Ultraviolet and Visible channel of the Wide Field Camera
3 (UVIS/WFC3) or the Wide Field Channel of the Advanced Camera for
Surveys (ACS/WFC). We then performed point-spread-function (PSF)
photometry on the calibrated scientific image frames (with extension
`flt') using the WFC3 or ACS modules of the DOLPHOT 2.0 package
\citep{Dolp16a}. After having obtained the raw stellar catalog, we
employed the same method as \cite{Li18c} to filter out `bad'
detections and compile the `clean' catalogs. Here, bad detections are
objects which were identified as bad pixels, cosmic rays, extended
sources, or centrally saturated objects. Objects with unrealistic
sharpness or which were strongly affected by their nearby crowded
environment were removed as well \cite[for details, see][]{Li18c}. In
this paper, we only focus on BSSs and normal evolved stars, such as
red-giant branch (RGB) stars, asymptotic giant branch (AGB) stars and
red clump (RC) stars. All these stars have high signal-to-noise ratios
(SNR$>$250); only central saturation will reduce their completeness
levels. We confirm that fewer than 5\% of the stars of interest would
be removed by our data reduction procedures.

In this paper, we aim to examine if the framework developed by
\cite{Ferr12a} and \cite{Lanz16a} for BSSs in old Galactic GCs could
also apply to younger GCs (younger than $\sim$10 Gyr). Therefore, we
do not perform photometry on clusters older than 10 Gyr. For clusters
younger than 10 Gyr, we only selected clusters containing more than 10
BSS candidates for reasons of statistical robustness. These selection
criteria forced us to exclude some young clusters, because those
clusters do not show a clear turnoff region and their BSSs evolve too
fast to be captured in an observational snapshot.

It is possible that some BSS candidates are just field stars, which
may occasionally be located along the line-of-sight direction. To
obtain reliable results about the radial distributions of BSSs,
field-star decontamination is required. However, because of the large
distances to the LMC clusters \citep[e.g.,][]{grijs14a}, using proper
motions to reduce pollution by field stars is not
possible\footnote{Most BSSs in our LMC clusters have magnitudes
  between 17 and 20 mag in the {\sl Gaia} $G$ band. The typical proper
  motion uncertainty for these stars is 0.2--1.2 mas yr$^{-1}$ (see
  \url{https://www.cosmos.esa.int/web/gaia/dr2}). At the distance of
  the LMC, $\sim 50$ kpc, this corresponds to $\sim$50--300 km
  s$^{-1}$, which is much larger than the central velocity dispersion
  of an LMC star cluster \citep[$<$5 km s$^{-1}$][]{Mcla05a}.}.
Therefore, we have visually inspected the spatial distributions of the
selected BSSs for each cluster. If these stars did not show any clear
central concentration, we did not analyze the cluster. Finally, only
seven clusters were selected for further study. Table \ref{T1} lists
the basic observational information for our seven sample clusters, as
well as that of their associated parallel observations centered on
nearby regions (if any), which are used as reference fields.

\begin{table*}
  \begin{center}
\caption{Description of the observations used. Clusters are ranked in
  terms of increasing best-fitting isochronal ages (see Table
  \ref{T2})}\label{T1}
  \begin{tabular}{l | l l l l l l}\hline
    Cluster      &  Camera	  & Exposure time & Filter & Program ID & PI name \\\hline
    NGC 1831 (Cluster \& Field)	& UVIS/WFC3 & 2$\times$975 s + 2$\times$1115 s	 & F336W	& GO-14688 & P. Goudfrooij & \\
    			&  		    & 100 s	 + 660 s + 720 s & F814W &  		  & 		       & \\\hline
    NGC 1868 (Cluster \& Field)	& UVIS/WFC3 & 2$\times$830 s + 831 s	 & F336W	& GO-14710 & A.~P. Milone & \\
    			&  		    & 90 s + 666 s & F814W &  		  & 		       & \\\hline
    NGC 2173 (Cluster)	& UVIS/WFC3 & 120 s + 2$\times$700 s & F475W	& GO-12275	& L. Girardi & \\
			&			&  30 s + 550 s + 2$\times$700 s  & F814W &	    		 &		       & \\
    NGC 2173 (Field)	& ACS/WFC & 90 s + 2$\times$500 s + 2$\times$700 s & F475W	& GO-12257	& L. Girardi & \\
			&			&  10 s + 600 s + 690 s + 2$\times$700 s & F814W &	    		 &		       & \\\hline
    NGC 2203 (Cluster)	& UVIS/WFC3 & 120 s + 2$\times$700 s & F475W	& GO-12275	& L. Girardi & \\
			&			&  30 s + 550 s + 2$\times$700 s  & F814W &	    		 &		       & \\
    NGC 2203 (Field)	& ACS/WFC & 90 s + 2$\times$500 s + 2$\times$700 s & F475W	& GO-12257	& L. Girardi & \\
			&			&  10 s + 550 s + 690 s + 2$\times$713 s & F814W &	    		 &		       & \\\hline
    NGC 2213 (Cluster)	& UVIS/WFC3 & 120 s + 600 s + 720 s & F475W	& GO-12275	& L. Girardi & \\
			&			&  30 s + 2$\times$700 s  & F814W &	    		 &		       & \\
    NGC 2213 (Field)	& ACS/WFC &  2$\times$500 s & F475W	& GO-12257	& L. Girardi & \\
			&			&  2$\times$500 s & F814W &	    		 &		       & \\\hline
    NGC 1651 (Cluster)	& UVIS/WFC3 & 120 s + 600 s + 720 s & F475W	& GO-12275	& L. Girardi & \\
			&			&  30 s + 550 s + 2$\times$700 s  & F814W &	    		 &		       & \\
    NGC 1651 (Field)	& ACS/WFC & 90 s + 2$\times$500 s + 2$\times$700 s & F475W	& GO-12257	& L. Girardi & \\
			&			&  30 s + 2$\times$700 s & F814W &	    		 &		       & \\\hline
    ESO 121-SC03 (Cluster \& Field)	& ACS/WFC & 90 s + 3$\times$360 s & F435W	& GO-10595	& P. Goudfrooij & \\
			&			&  8 s + 2$\times$350 s  & F814W &	    		 &		       & \\\hline			
  \end{tabular} 
  \end{center} 
\end{table*} 

\subsection{Selection of the stellar samples}

\begin{figure*}
\includegraphics[width=2\columnwidth]{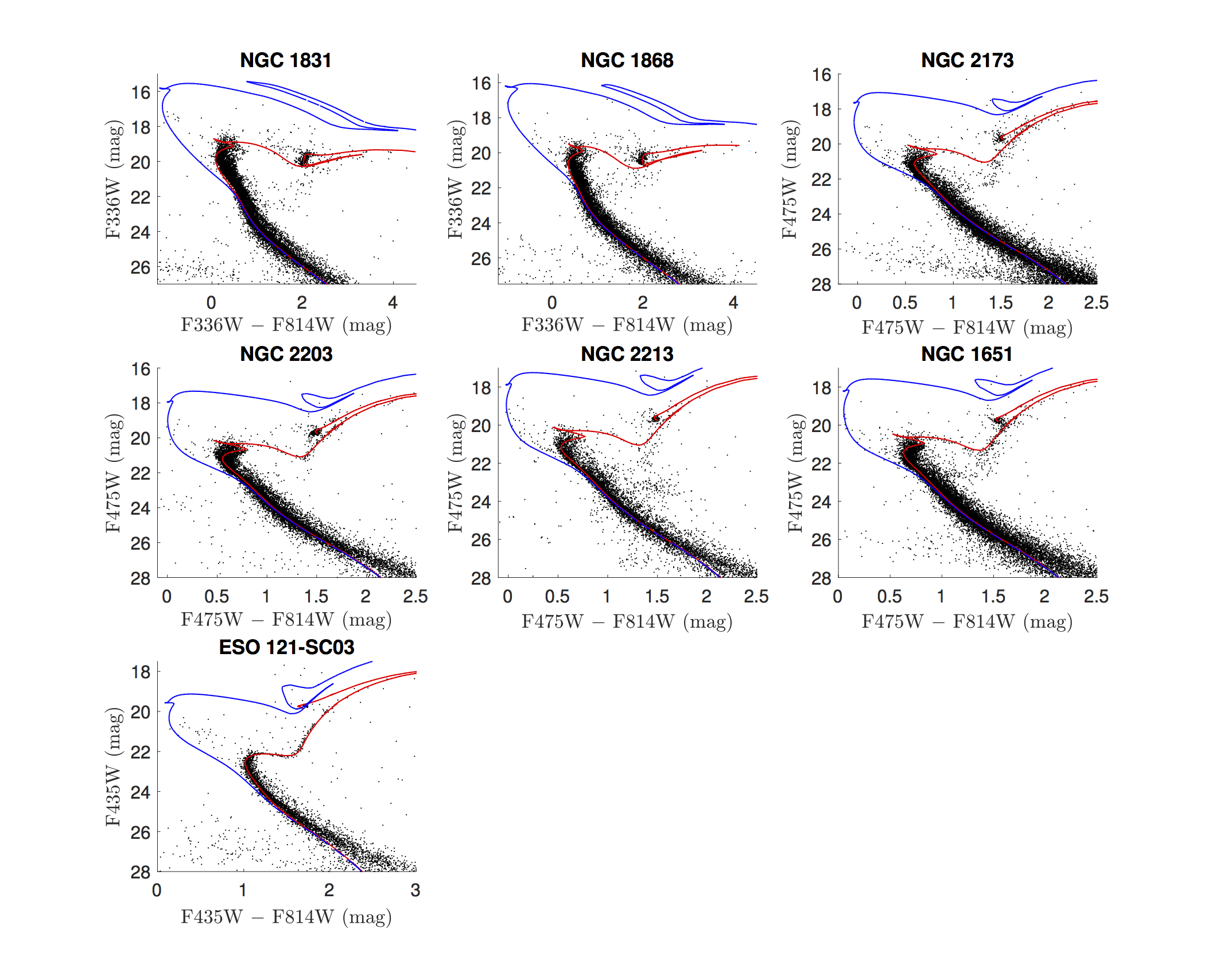}
\caption{Processed CMDs for all clusters. Red solid lines are
  isochrones describing the bulk stellar populations of the star
  clusters. Blue dashed lines are the corresponding young isochrones
  with turnoff stellar masses twice those of the old isochrones.}
\label{F1}
\end{figure*}

In Fig. \ref{F1} we present the processed CMDs of our clusters, along
with their best-fitting isochrones calculated based on the PARSEC
stellar evolution code \citep{Bres12a}. For each cluster, we use an
old isochrone to fit the bulk stellar population based on visual
inspection. Most clusters exhibit an extended main-sequence turnoff
(eMSTO) region, a feature that is commonly found in clusters younger
than $\sim$1--2 Gyr \citep[e.g.,][]{Milo09a,Mari18a}. The eMSTO makes
determining an exact age for the bulk stellar population
difficult. However, some intermediate-age clusters with eMSTO regions
exhibit tight subgiant branches \citep[SGBs;][]{Li14a,Bast15a}.
Therefore, we adopt the isochrone that approximately describes the SGB
ridge-line as the best-fitting isochrone. If a cluster does not show a
well-populated SGB, we adopt an isochrone that describes the blue
boundary of the turnoff region as the best-fitting isochrone. To
search for a cluster's BSSs, we adopted another young isochrone
characterized by a turnoff stellar mass twice that of the old
isochrone. Both isochrones were calculated for the same metallicity,
extinction, and distance modulus; their only difference are their
respective ages. Parameters for the adopted isochrones are presented
in Table \ref{F2}, where we have included the mass of the old
isochrones' turnoff stars as well. NGC 2213 was recently studied by
\cite{Li18a}; we directly apply these authors' best-fitting isochrones
to fit its CMD.

\begin{table*}
  \begin{center}
\caption{Basic model fit parameters.}\label{T2}
  \begin{tabular}{c | c c c c c c}\hline
    Cluster      &  log $t_{\rm 1}$ & log $t_{\rm 2}$ & $Z^*$& $A_V$& $(m-M)_{0}$ & $M_{\rm TO}$\\
    	&	[yr]	&	[yr]	&    & (mag) & (mag) & ($M_{\odot}$)\\
	& (1) & (2) & (3) & (4) & (5) & (6)\\\hline
    NGC 1831 	& 8.85 & 8.05 & 0.006 & 0.12 & 18.50 & 2.21 \\
    NGC 1868 	& 9.08 & 8.20 & 0.006 & 0.06 & 18.50 & 1.82 \\
    NGC 2173 	& 9.25 & 8.45 & 0.008 & 0.16 & 18.45 & 1.62 \\
    NGC 2203 	& 9.25 & 8.49 & 0.006 & 0.22 & 18.50 & 1.59 \\
    NGC 2213 	& 9.26 & 8.48 & 0.006 & 0.06 & 18.50 & 1.59 \\
    NGC 1651 	& 9.30 & 8.52 & 0.005 & 0.30 & 18.55 & 1.50 \\
    ESO 121-SC03 & 9.84 & 9.00 & 0.001 & 0.27 & 18.40 & 0.94\\\hline
  \end{tabular} 
  \end{center} 
  (1) Age of the bulk population stars (in logarithmic units). (2) Age
  of the young isochrone (3) Metallicity. (4) Reddening. (5) Distance
  modulus. (6) Turnoff stellar mass for the bulk population\\ $*$:
  $Z_{\odot}=0.0152$.
\end{table*} 

\subsubsection{Blue straggler stars}

The next step involved selecting BSSs and their reference stellar
populations, which is similar to \cite{Lanz16a}'s approach to study
their Galactic GCs. For most clusters, the reference stellar
populations are composed of RGB, RC, and AGB stars. Since the RGB and
AGB are poorly populated in NGC 1831 and NGC 1868, we only selected
the populations of RC stars as these clusters' reference
populations. Our BSS selection approach proceeded as follows.

\begin{enumerate}

\item We shifted each cluster's old isochrone to run across the region
  where the stellar number density is significantly lower than that in
  the MSTO region. Specifically, the position of this boundary was
  determined as follows. We first determined the best-fitting old
  isochrone describing the red boundary of the BSS region. We then
  checked how many stars would be selected as BSS candidates. We then
  shifted this isochrone adopting color steps of $\sim$0.01 mag to the
  blue. The number of selected BSS candidates would initially decrease
  rapidly, because many MSTO stars were removed. Once the rate of the
  decrease dropped below 5\% per 0.01 mag, we defined the shifted
  isochrone as the boundary separating BSSs from MSTO stars. We
  emphasize that this method was adopted to minimize the number
  uncertainties. In reality, one would not expect a sharp boundary
  between BSSs and MSTO stars, and there are probably collision
  products and binary mass transfer products mixed in with MS stars.

\item We defined stars located along the extension of the MS that were
  bluer than the red boundary of the MSTO region but redder than the
  young isochrone combined with a negative photometric color spread
  \citep[which was determined to take the photometric uncertainties
    and possible differential reddening into account, see][]{Li18c} as
  our BSS sample.

\item We also adopted a lower magnitude limit for our BSSs. To
  avoid contamination by MS stars, we adopted the locus where the
  young and old isochrones begin to diverge significantly (their color 
  separations are greater than at least 0.1 mag) as our lower
  boundary.

\item We do not set any upper boundary for our BSSs, because we found
  that the adopted young isochrone covers all observed stars brighter
  than the MSTO region.
\end{enumerate}

\subsubsection{Reference population stars}

\begin{figure*}
\includegraphics[width=2\columnwidth]{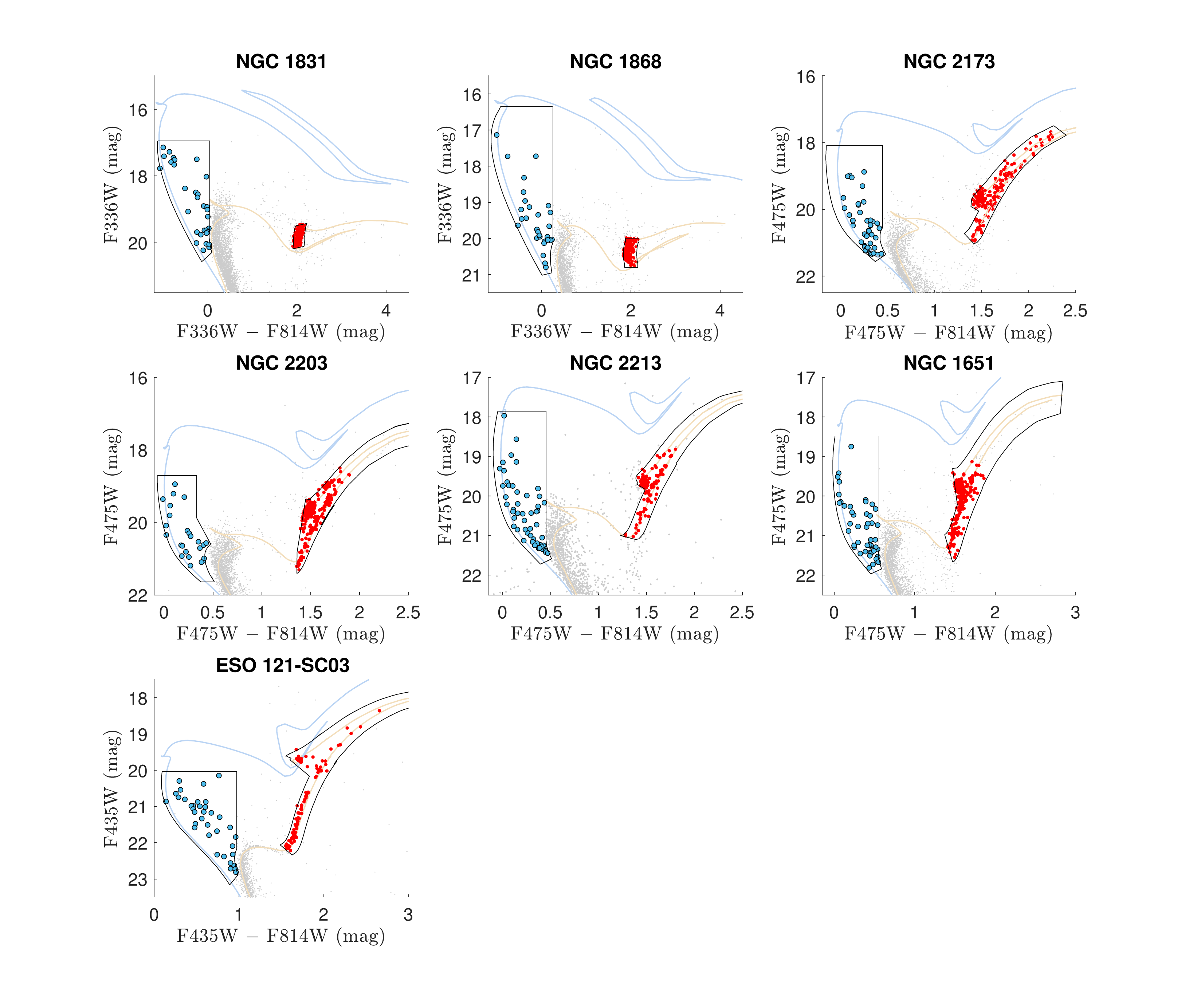}
\caption{Illustration of how we select BSSs (blue circles) and
  reference population stars (RGB, RC, and AGB stars; indicated by red
  dots).}
\label{F2}
\end{figure*}

For NGC 1831 and NGC 1868, we only selected RC stars as reference
stellar populations, because there are no well-populated RGB or AGB
features in their CMDs (see Fig. \ref{F1}). For these two clusters, we
simply defined a box which approximately covered the bulk of their RC
stars as the typical RC region. For the other clusters, a combination
of RGB, RC, and AGB stars was selected as reference population. This
is similar to the approach adopted by \cite{Lanz16a}, who defined
stars on the RGB, SGB, and/or horizontal branch as reference
populations. To select reference populations in our other clusters, we
used the CMD section following the bottom of the RGB of the old
isochrone as the ridge-line. We then applied a large color--magnitude
spread to define the relevant boundary; the typical magnitude
deviation from the ridge-line was about 0.1--0.3 mag, which we adopted
so as to consider the possible scatter caused by photometric errors
and differential reddening. We illustrate our selection method in
Fig. \ref{F2}. We also assumed that there are number uncertainties
associated with the normal distributions defining these stellar
populations.

\subsection{Structural parameters}

\begin{figure}
\includegraphics[width=\columnwidth]{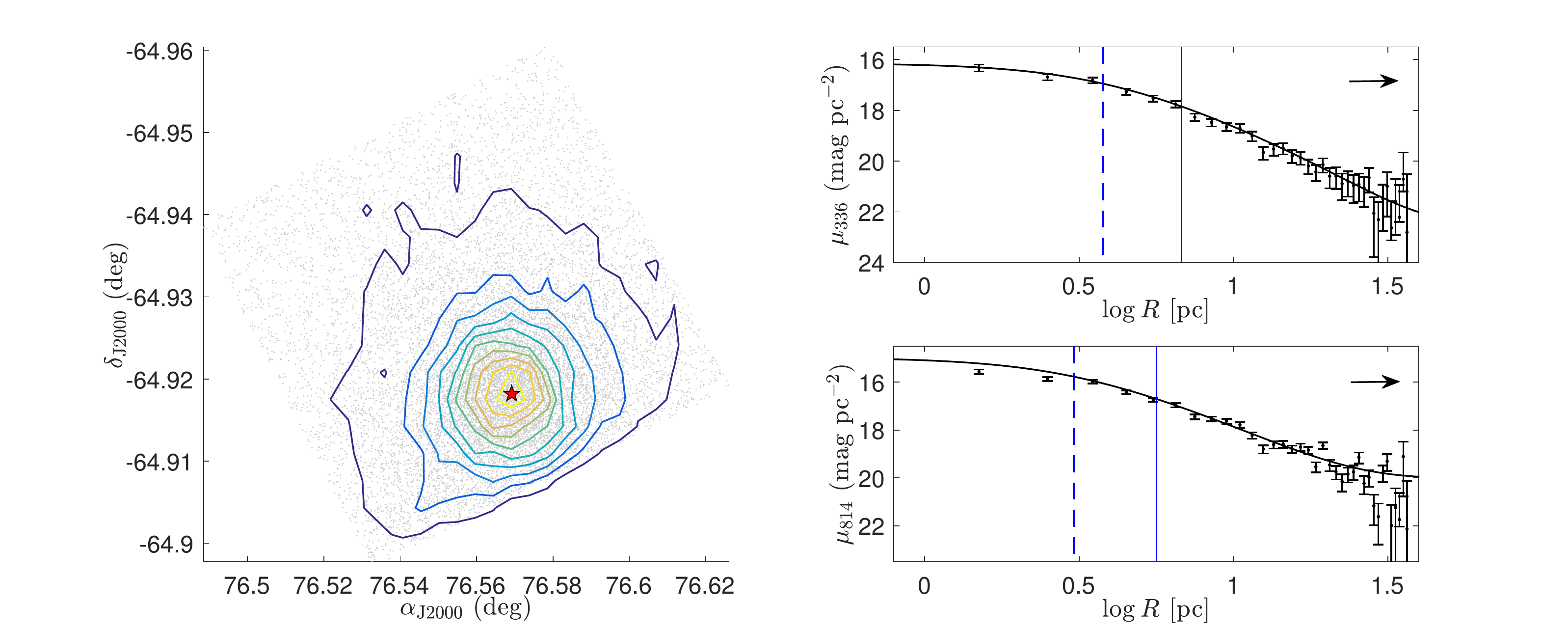}
\caption{(left) Stellar spatial distribution and number density
  contours for NGC 1831. The red pentagram is the calculated cluster
  center. (right) Brightness profile of NGC 1831 in two passbands
  (top: F336W; bottom: F814W). Blue dashed and solid lines indicate
  the best-fitting core and half-light radii. Black arrows mean that
  the best-fitting tidal radii associated with these profiles are
  beyond the figure boundaries.}
\label{F3}
\end{figure}

\begin{figure}
\includegraphics[width=\columnwidth]{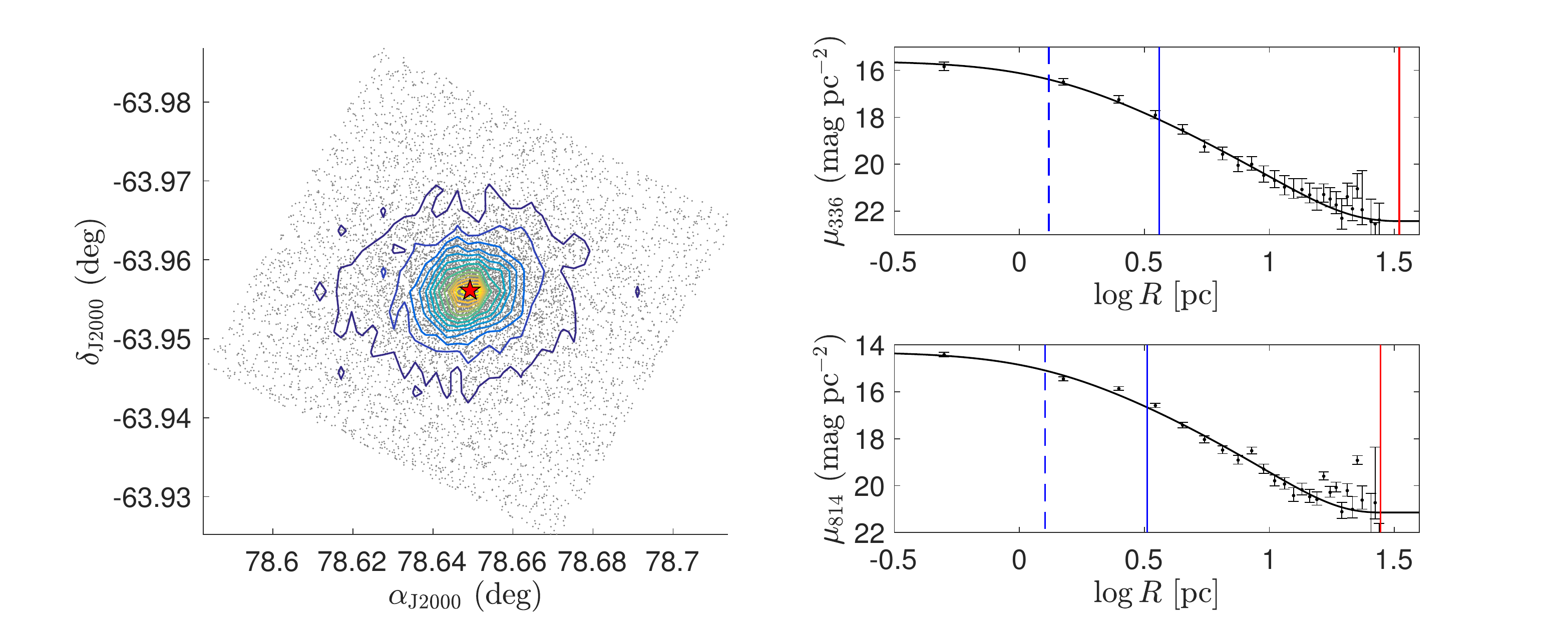}
\caption{As Fig. \ref{F3}, but for NGC 1868. In the right-hand panels,
  red solid lines indicate the best-fitting tidal radii in both
  passbands.}
\label{F4}
\end{figure}

\begin{figure}
\includegraphics[width=\columnwidth]{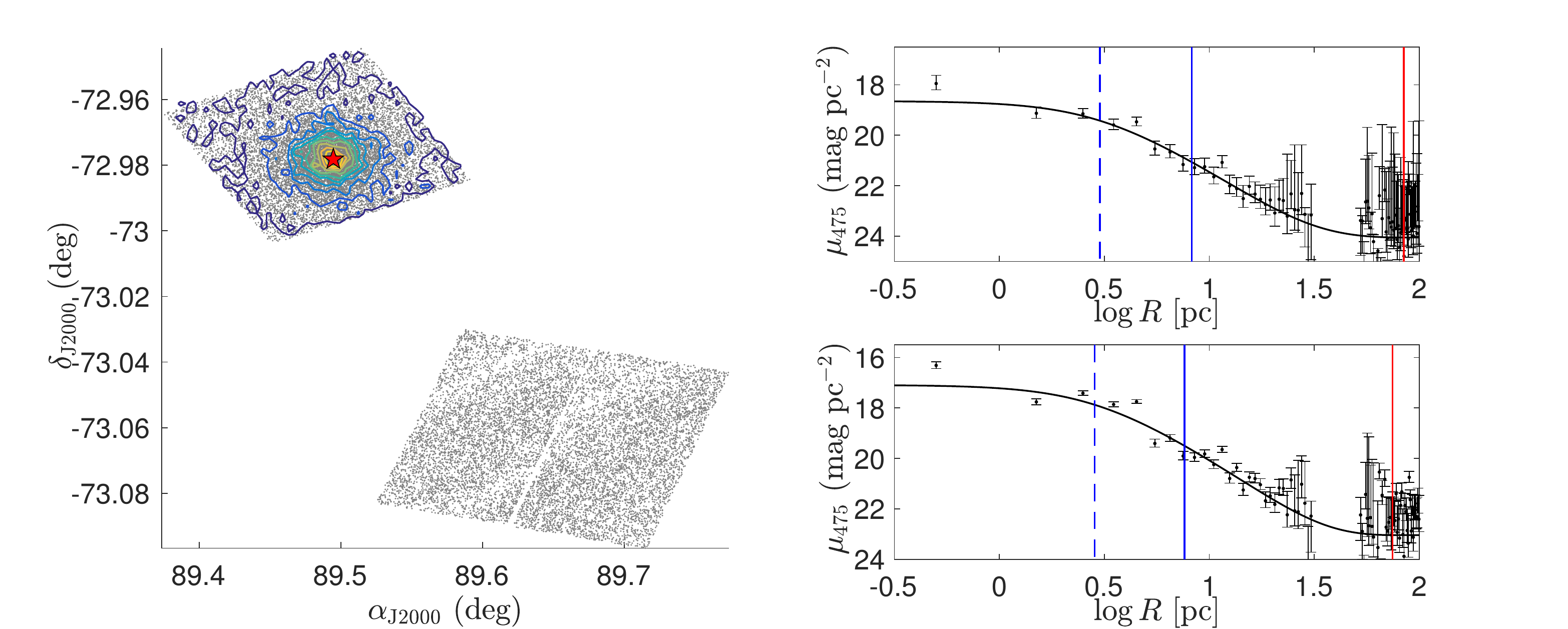}
\caption{As Fig. \ref{F4}, but for NGC 2173. The bottom square in the
  left-hand panel represents a nearby field observation.}
\label{F5}
\end{figure}

\begin{figure}
\includegraphics[width=\columnwidth]{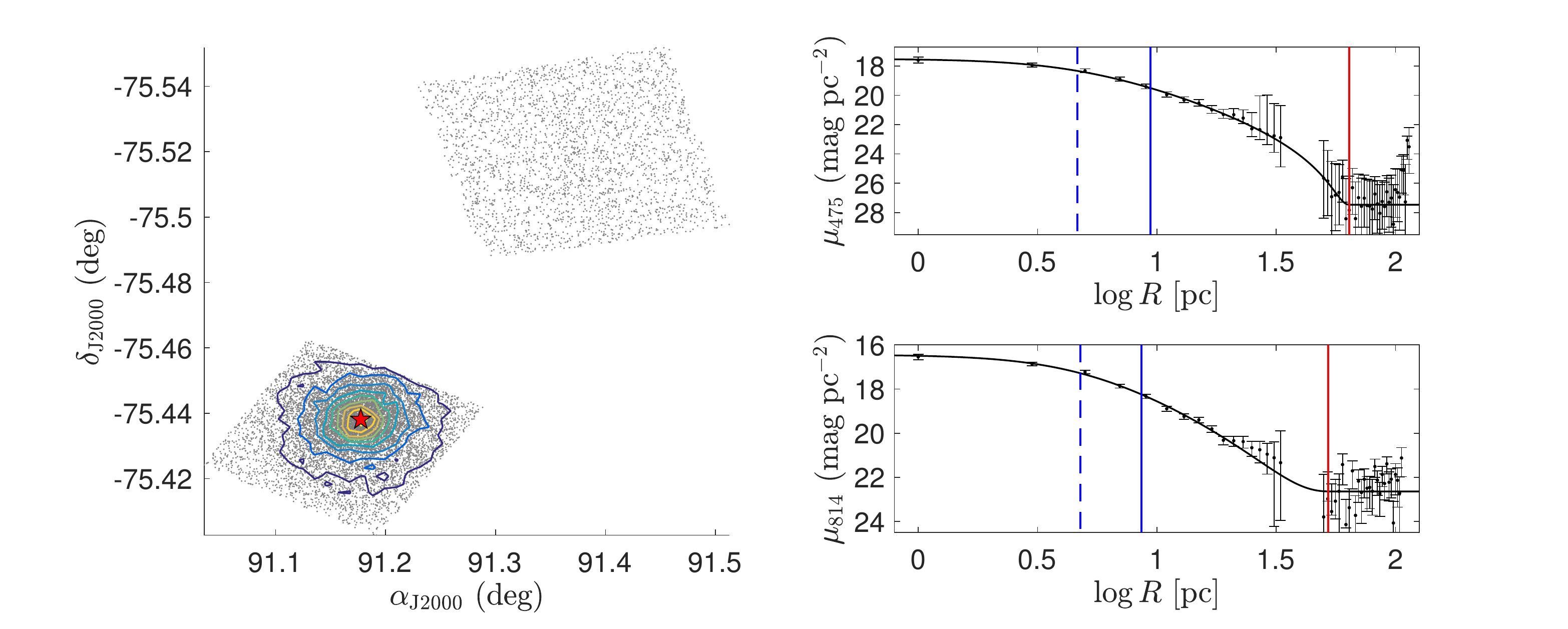}
\caption{As Fig. \ref{F5}, but for NGC 2203.}
\label{F6}
\end{figure}

\begin{figure}
\includegraphics[width=\columnwidth]{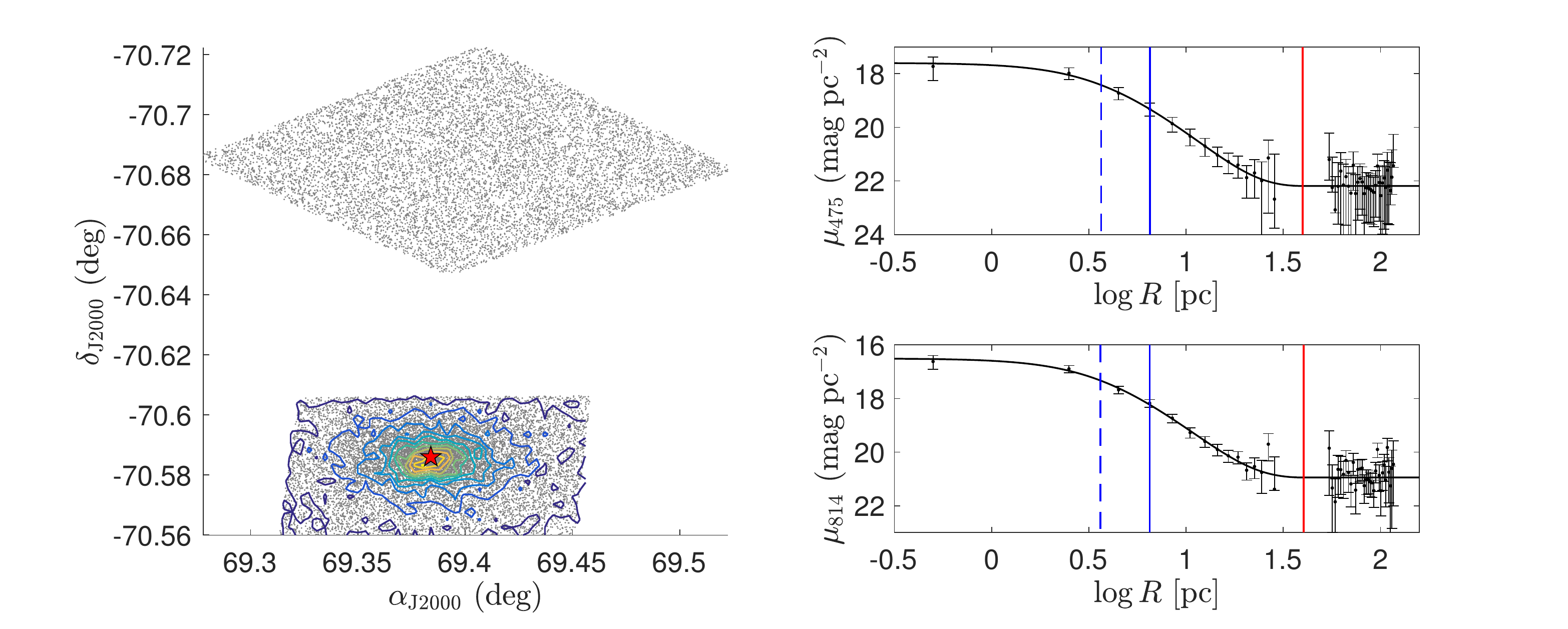}
\caption{As Fig. \ref{F5}, but for NGC 1651.}
\label{F7}
\end{figure}

\begin{figure}
\includegraphics[width=\columnwidth]{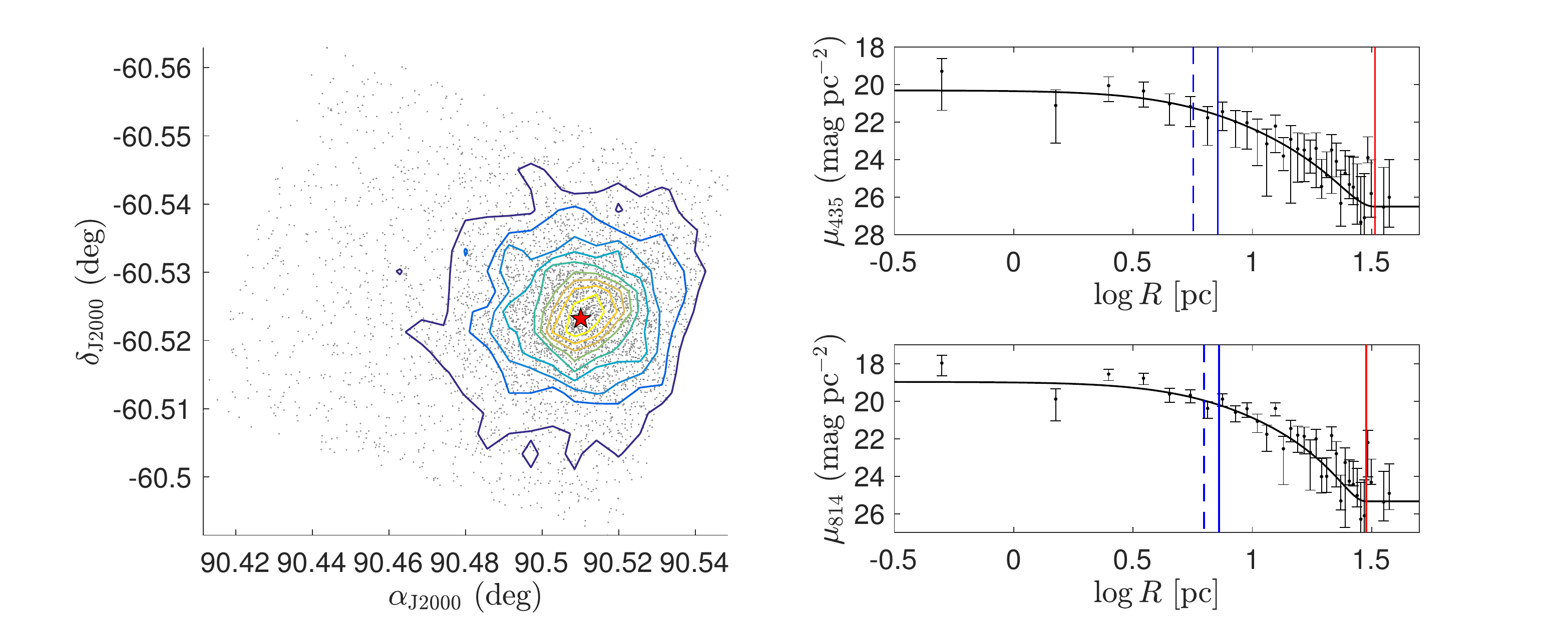}
\caption{As Fig. \ref{F3}, but for ESO 121-SC03.}
\label{F8}
\end{figure}

We calculated the center coordinates of our clusters using the method
described by \cite{Li18b}. We calculated the stellar number density
contours for the stars detected in the field of view (FoV) of our
observations. The position where the stellar number density reaches
its maximum value was defined as the cluster center. The stellar
spatial distributions of our clusters, as well as their number density
contours and the derived cluster centers, are presented in the
left-hand panels of Figs \ref{F3}--\ref{F8}. For NGC 2213, we used the
cluster center determined by \cite{Li18a}. We calculated the clusters'
brightness profiles in two passbands. Because massive stars are
expected to contribute most of the flux, in each passband we only
selected stars brighter than a given magnitude to calculate their
brightness profiles. This magnitude limit varied from cluster to
cluster; it was usually two or three magnitudes brighter than the
detection limit. We only selected these bright stars because, they are
all characterized by high completeness levels. We calculated their
brightness profiles using the method adopted by \cite{Li18a}. If a
cluster had a seperate field observation, we also extended their
brightness profile to the field region, adopting a constant brightness
level. Using least-squares minimization, we used a King model to fit
the calculated brightness profile \citep{King62a}:
\begin{equation}
\mu(r)=k\left[\frac{1}{\sqrt{1+(r/r_{\rm
        c})^2}}-\frac{1}{\sqrt{1+(r_{\rm t}/r_{\rm c})^2}}\right]+{b}.
\end{equation}
Here, $r_{\rm c}$ and $r_{\rm t}$ are the core and tidal radii,
respectively, $b$ is a constant representing the background
brightness, and $k$ is a normalization coefficient. The calculated
brightness profiles as well as the best-fitting King models are shown
in the right-hand panels of Figs \ref{F3}--\ref{F8} \citep[for NGC
  2213, see][their Fig. 3]{Li18a}.

Based on the best-fitting King models, we also determined the
best-fitting half-light radii in both passbands. These derived
structural parameters are presented in Table \ref{T3}. For most of our
clusters, we found that the derived structural parameters are
consistent within the uncertainties. We thus selected their average
values as the best-fitting structural parameters.


\begin{table*}
  \begin{center}
\caption{Best-fitting structural parameters based on King profile
  fits. (1) Cluster name, (2) right ascension, (3) declination, (4)
  best-fitting core radius in band A, (5) best-fitting half-light
  radius in band A, (6) best-fitting tidal radius in band A, (7)
  best-fitting core radius in band B, (8) best-fitting half-light
  radius in band B, (9) best-fitting tidal radius in band B. Here,
  `band A' is either F336W, F435W, or F475W, while `band B' is
  F814W. Problematic values or uncertainties are indicated in bold
  font; they were not used. The structural parameters of NGC 2213 were
  derived by \cite{Li18a} (only calculated in the F475W
  passband).}\label{T3}
  \begin{tabular}{c | c c | c c c | c c c}\hline
    Cluster      &  $\alpha_{\rm J2000}$ & $\delta_{\rm J2000}$ & $r_{\rm 1c}$ (pc) & $r_{\rm 1hl}$ (pc) & $r_{\rm 1t}$ (pc) & $r_{\rm 2c}$  (pc) & $r_{\rm 2hl}$ (pc) & $r_{\rm 2t}$ (pc) \\
   (1) & (2) & (3) & (4) & (5) & (6) & (7) & (8) & (9)\\\hline
    NGC 1831 & 05$^{\rm h}$06$^{\rm m}$16.56$^{\rm s}$ & $-$64$^{\circ}$55$'$05.52$''$ & 3.78$\pm$0.22 & 6.79$^{+2.36}_{-3.67}$ & 60.86$\pm$45.20 & 3.04$\pm$0.21 & $5.64^{+1.37}_{-0.34}$ & $52.12\pm{24.79}$\\
    
    NGC 1868 & 05$^{\rm h}$14$^{\rm m}$35.88$^{\rm s}$ & $-$63$^{\circ}$57$'$21.96$''$ & 1.31$\pm$0.12 & 3.63$^{+0.31}_{-0.37}$ & 33.06$\pm$7.73 & 1.27$\pm$0.20 & 3.25$^{+0.45}_{-0.49}$ & 27.83$\pm$5.28\\
    
    NGC 2173 & 05$^{\rm h}$57$^{\rm m}$58.68$^{\rm s}$ & $-$72$^{\circ}$58$'$41.52$''$ & 3.01$\pm$0.40 & 8.53$^{+1.67}_{-2.53}$ & 84.27$\pm48.57$ & 2.84$\pm$0.44 & 7.61$^{+1.81}_{-2.13}$ & 74.56$\pm$43.74\\
    
    NGC 2203 & 06$^{\rm h}$04$^{\rm m}$42.60$^{\rm s}$ & $-$75$^{\circ}$26$'$17.16$''$ & 4.64$\pm$0.15 & 9.40$^{+0.30}_{-0.31}$ & 63.99$\pm$4.55 & 4.78$\pm$0.37 & 8.62$^{+0.59}_{-0.60}$ & 52.28$\pm$7.88\\
    
    NGC 2213 & 06$^{\rm h}$10$^{\rm m}$42.24$^{\rm s}$ & $-$71$^{\circ}$31$'$44.76$''$ & 1.45$\pm$0.02 & 3.59$^{+0.41}_{-0.59}$ & 31.41$\pm$3.02 & -- & -- & --\\
    
    NGC 1651 & 04$^{\rm h}$37$^{\rm m}$32.16$^{\rm s}$ & $-$70$^{\circ}$35$'$09.60$''$ & 3.66$\pm$0.30 & 6.52$^{+0.46}_{-0.45}$ & 39.78$\pm$5.40 & 3.63$\pm$0.42 & 6.48$^{+0.70}_{-0.72}$ & 40.26$\pm$8.11\\
    
    ESO 121-SC03 & 06$^{\rm h}$02$^{\rm m}$02.40$^{\rm s}$ & $-$60$^{\circ}$31$'$23.52$''$ & 5.66$\pm$1.14 & 7.18$^{+1.00}_{-1.01}$ & 32.64$\pm$7.73 & 6.28$\pm$1.50 & 7.25$^{+1.08}_{-1.04}$ & 30.05$\pm$6.42\\\hline
  \end{tabular} 
  \end{center} 
\end{table*} 

\begin{figure*}
\includegraphics[width=2\columnwidth]{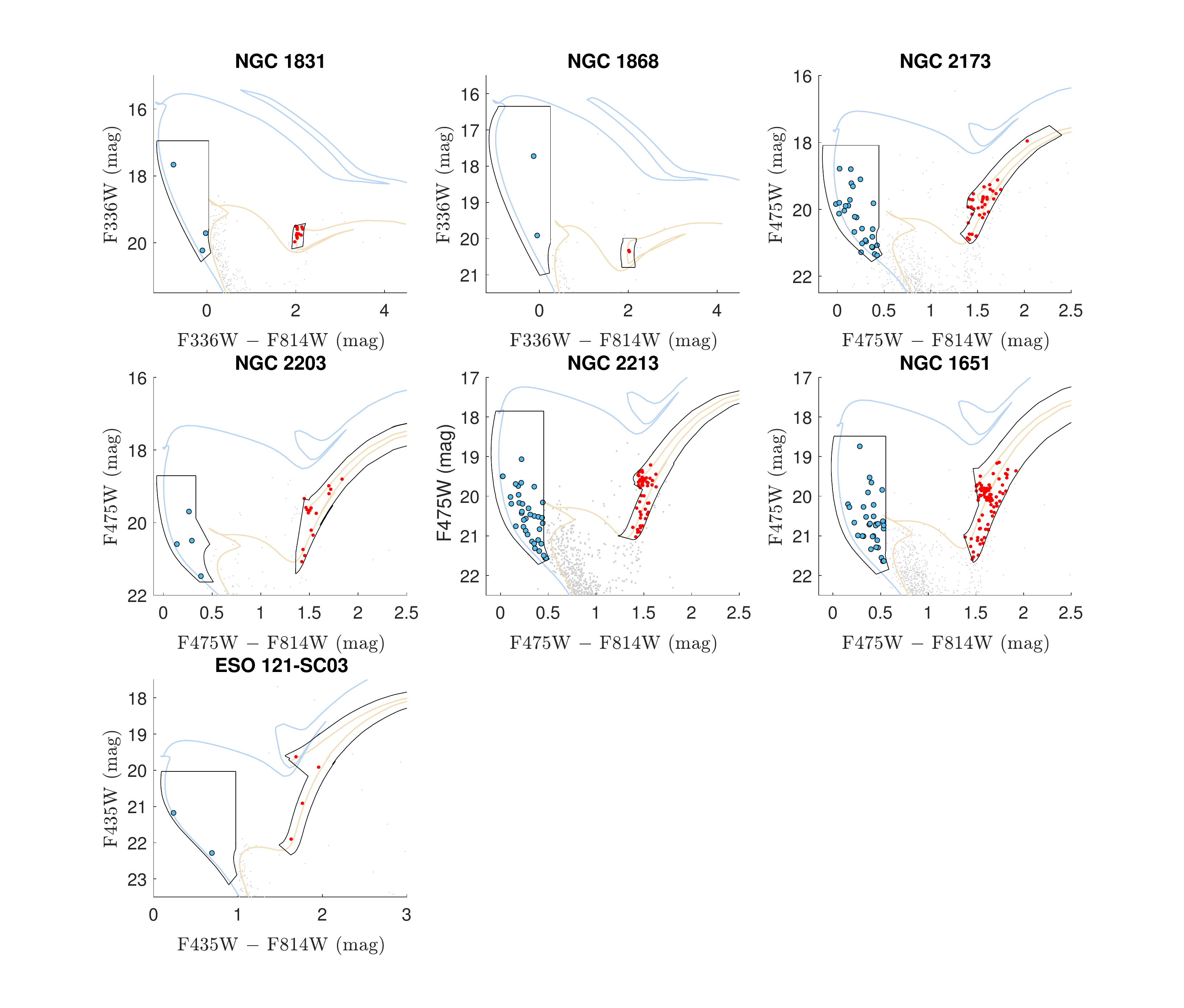}
\caption{As Fig. \ref{F2}, but for the observations of the
  corresponding reference fields. Note that the reference fields are
  larger than the cluster regions (see Table \ref{T4}).}
\label{F9}
\end{figure*}

\begin{figure*}
\includegraphics[width=2\columnwidth]{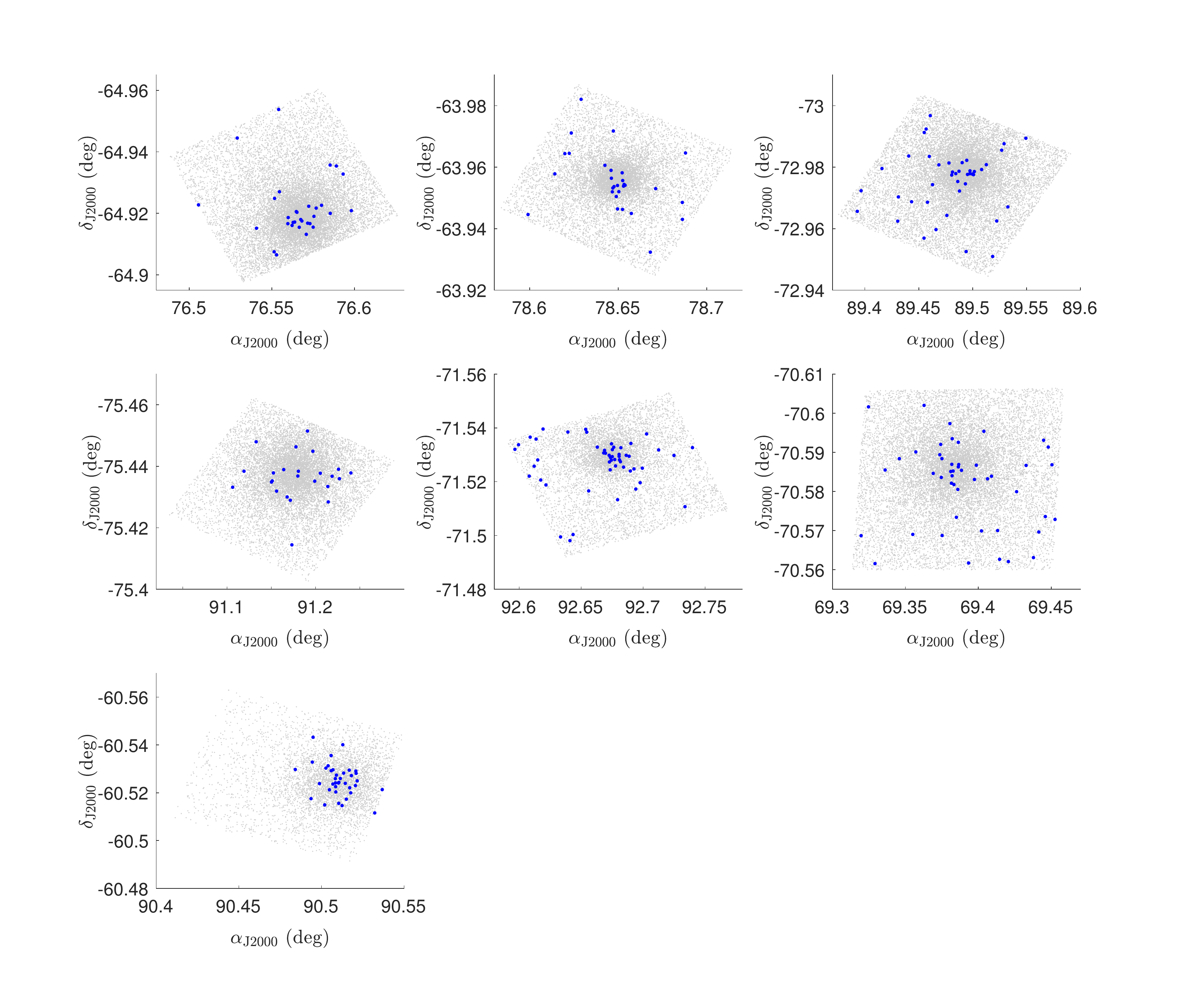}
\caption{BSS spatial distributions (blue dots) in our sample
  clusters.}
\label{F10}
\end{figure*}

In this paper, we only analyze stars within the half-light radius,
which we adopt as the cluster region. For NGC 1831, NGC 1868, and ESO
121-SC03, there are no separate observations for use as reference
field. We adopted the regions at radii greater than 20 pc as our
reference fields. As shown in Table \ref{T3}, these radii are too
small compared with the corresponding tidal radii. This means that we
must have overestimated the field contamination for NGC 1831, NGC
1868, and ESO 121-SC03. In principle, overestimation of the field
contamination would not affect the derived radial distributions of
their BSSs, because we have assumed a flat distribution of field stars
in our analysis. However, it will increase the associated
uncertainties when we calculate their central concentrations (see
Section \ref{S3}). For NGC 2173, NGC 2203, NGC 2213, and NGC 1651,
parallel observations located beyond the tidal radii were used as
reference fields. The selected field stars are shown in
Fig. \ref{F9}. However, this figure cannot reflect the actual field
contamination level, because the field areas are much larger than the
cluster regions. In Table \ref{T4}, we list the area ratios of the
reference fields and the cluster regions. To show that the derived
radial BSS distributions are not field artefacts, we also confirmed
that all BSSs in our clusters show apparent central concentrations
(see Fig. \ref{F10}), whereas the adopted field stars do not.

\subsection{Stellar completeness}

The next step is to correct for the differences in completeness
between the stellar samples in the cluster regions and the reference
fields. For NGC 1831, NGC 1868, and ESO 121-SC03, there are no
significant completeness differences, because their reference fields
and cluster regions were observed as part of the same image. However,
the reference fields associated with NGC 2173, NGC 2203, NGC 2213, and
NGC 1651 may be characterized by significantly different completeness
levels. Therefore, we have to calculate the completeness levels for
all stellar samples in both the cluster regions and the reference
fields. To do this, we generated 100,000 artificial stars with the
same color--magnitude distributions to the selected stellar
samples. We then repeated our {\sc Dolphot} approach in {\sl
  artificial star} mode 1,000 times. Each time we only photometered
100 artificial stars. We did not do the photometry for these 100,000
artificial stars at one time because adding too many stars to the raw
images would increase the crowding in the FoV, unnecessarily reducing
the stellar completeness. For clusters with a separate reference
field, we also applied the {\sl artificial star}-mode photometry to
the field observations. For the resulting artificial stars, {\sc
  dolphot} provides the same photometric parameters (magnitudes,
crowding, sharpness, etc.). We employed the same data reduction
process to the artificial stars as to the real observations, obtaining
a `clean' catalog of artificial stars (see Section \ref{S2a}). The
number ratio of the artificial stars in the resulting `clean' catalog
and in the raw catalog defines the stellar completeness: see Table
\ref{T4}.

\begin{table}
  \begin{center}
\hspace{-1cm}\caption{Completeness for different stellar samples and
  field-to-cluster area ratios. (1) Cluster name, (2) completeness of
  BSSs in the cluster area, (3) completeness of reference population
  stars (RGB, RC, and AGB stars) in the cluster area, (4) completeness
  of BSSs in the reference field, (5) completeness of reference
  population stars in the reference field, (6) area ratio of the
  reference field to the adopted cluster region.}\label{T4}
  \begin{tabular}{c | c c | c c | r}\hline
    Cluster      &  $f_{\rm b}$ & $f_{\rm r}$ & $f_{\rm bf}$ & $f_{\rm rf}$ & $A_{\rm f}/A_{\rm c}$ \\
   (1) & (2) & (3) & (4) & (5) & (6) \\\hline
   NGC 1831 & 99\% & 99\% & -- & -- & 4.7 \\
   NGC 1868 & 99\% & 99\% & -- & -- & 7.8 \\
   NGC 2173 & 99\% & 98\% & $\sim$100\% & 97\% & 11.8 \\
   NGC 2203 & $\sim$100\% & 49\% & $\sim$100\% & 96\% & 9.3 \\
   NGC 2213 & 99\% & 95\% & 99\% & 43\% & 56.3 \\
   NGC 1651 & $\sim$100\% & 98\% & 99\% & 37\% & 18.3 \\
   ESO 121-SC03 & 99\% & 97\% & -- & -- & 7.7 \\
    \hline
  \end{tabular} 
  \end{center} 
\end{table} 

\begin{figure}
\includegraphics[width=\columnwidth]{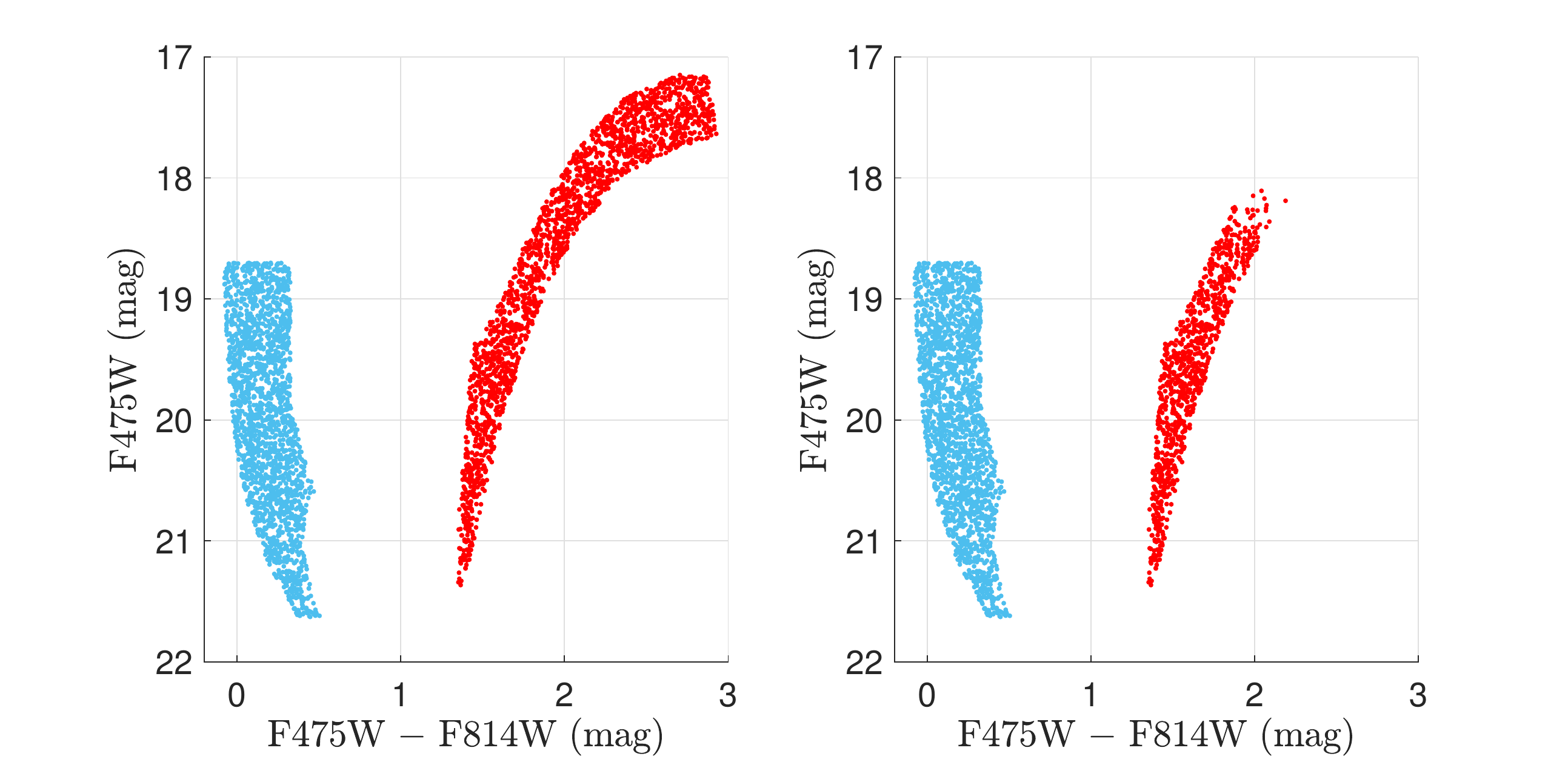}
\caption{Input (left) and output (right) CMDs for artificial stars
  corresponding to NGC 2203. Blue and red dots represent the
  artificial BSSs and normal evolved stars (reference population
  stars). As shown in the right-hand panel, most normal evolved stars
  located in the upper RGB are saturated.}
\label{F11}
\end{figure}

The completeness of most stellar samples in our clusters is almost
unity. This is expected, since these stars have high SNRs ($>$250). By
examining the sharpness distribution of these stars, we confirmed that
they are unlikely extended sources (e.g., faint background galaxies)
or sharp sources (such as cosmic rays). However, for NGC 2203 we found
that the stellar completeness in the region of the reference
population is only 49\%. This is because some bright objects, such as
the thermally pulsing AGB stars, are saturated. This also applies to
the reference population stars in the reference fields of NGC 2213 and
NGC 1651, which yield average completeness levels of 43\% and 37\%,
respectively. By exploring the characteristic of the artificial stars,
we confirmed that this is because some stars in the F814W passband are
saturated. If we constrain the reference population stars to
F814W$\geq$16.5 mag for NGC 2203, and F814W$\geq$17.0 mag and
F814W$\geq$17.4 mag for the reference fields of NGC 2213 and NGC 1651,
respectively, their overall completeness will be greater than 95\%. As
a consequence, for NGC 2203, NGC 2213, and NGC 1651, we only selected
normal evolved stars fainter than these magnitude limits. This is why
in Figs \ref{F2} and \ref{F9} the reference population stars seem to
be cut off in magnitude. In Fig. \ref{F11} we show an example of the
input (left-hand panel) and output CMDs for artificial stars generated
for NGC 2203.

\section{Main Results}\label{S3}

In this section we study the radial distributions of the BSSs in
  our sample clusters. \cite{Ales16a} and \cite{Lanz16a} defined a
parameter $A^{+}_{\rm rh}$, which represents the area enclosed between
the cumulative radial distributions of the BSSs and the reference
samples. Following their framework, we calculated $A^{+}_{\rm rh}$ for
the BSSs in our clusters. Specifically, $A^{+}_{\rm rh}$ is defined as
\begin{equation}
A^{+}_{\rm rh}=\int_{-\infty}^{0}{\phi_{\rm b}(x)-\phi_{\rm r}(x){\rm d}x},
\end{equation}
where $x=\log{(r/r_{\rm hl})}$ is the logarithm of the radius from the
cluster center normalized to the half-light radius $r_{\rm hl}$;
$x=-\infty$ and $x=0=\log{1}$ correspond to $r=0$ and $r=r_{\rm hl}$,
$\phi_{\rm b}(x)$ and $\phi_{\rm r}(x)$ are the cumulative radial
distributions of BSSs and reference population stars. A large value of
$A^{+}_{\rm rh}$ means that the BSSs are more centrally concentrated
than the reference population stars. If $A^{+}_{\rm rh}$ is equal to
zero or even negative, this would indicate that the BSSs are fully
mixed with the reference population stars, or even less segregated.

The uncertainty adopted for our calculation includes two
components. The first comes from the normal distribution of the
uncertainties in the stellar total number. The second part is caused
by the unknown field contamination. For clusters beyond the Milky Way,
only statistical methods are applicable to estimate the field
contamination. This method compares the CMD of the cluster region with
that of its reference field and statistically subtracts stars from the
cluster CMD by assuming that field stars will have the same
color--magnitude distribution in the cluster region. This method has
been widely used for LMC clusters \citep[e.g.,][]{Milo18a,Yang18a}.
However, it cannot be directly applied in this paper, because (1) both
the numbers of genuine cluster members and contaminating field stars
are very small in the BSS regions. The small number dispersion will
lead to large uncertainties in the number ratios in different radial
bins. (2) We aim to study the radial distribution of BSSs. Randomly
subtracting a small number of field stars without careful
consideration of their spatial positions may significantly change the
result, thus reducing the reliability of our conclusion.

Because of these limitations, we devised a Monte-Carlo based method to
estimate the `average' number fraction radial profiles. Before
applying this method, we first confirmed that the total number of BSSs
in the cluster regions cannot be fully explained by field
contamination. In Table \ref{T5} we list the observed number of BSSs
and reference population stars in the cluster regions and the expected
numbers of contaminating field stars (after correction for the area
difference between cluster regions and the reference fields).

\begin{table}
  \begin{center}
\caption{Observed numbers of (2) BSSs and (3) reference population
  stars in the cluster regions, and the expected numbers of
  contaminating field stars in the regions of (4) BSSs and (5)
  reference population stars.}\label{T5}
  \begin{tabular}{c | c c | c r}\hline
    Cluster      &  $N_{\rm b}$ & $N_{\rm r}$ & $N_{\rm bf}$ & $N_{\rm rf}$  \\
   (1) & (2) & (3) & (4) & (5)   \\\hline
   NGC 1831 & 19 & 107 & 1--2 & $\sim$3 \\
   NGC 1868 & 11 & 84 & $\sim$0 & $\sim$0 \\
   NGC 2173 & 19 & 126 & $\sim$2 & $\sim$3 \\
   NGC 2203 & 13 & 242 & 0--1 & 1--2 \\
   NGC 2213 & 18 & 61 & 0--1 & $\sim$1 \\
   NGC 1651 & 19 & 122 & 1--2 & 4--5 \\
   ESO 121-SC03 & 27 & 52 & $\sim$0 & 0--1 \\
    \hline
  \end{tabular} 
  \end{center} 
\end{table} 

\begin{figure*}
\includegraphics[width=2\columnwidth]{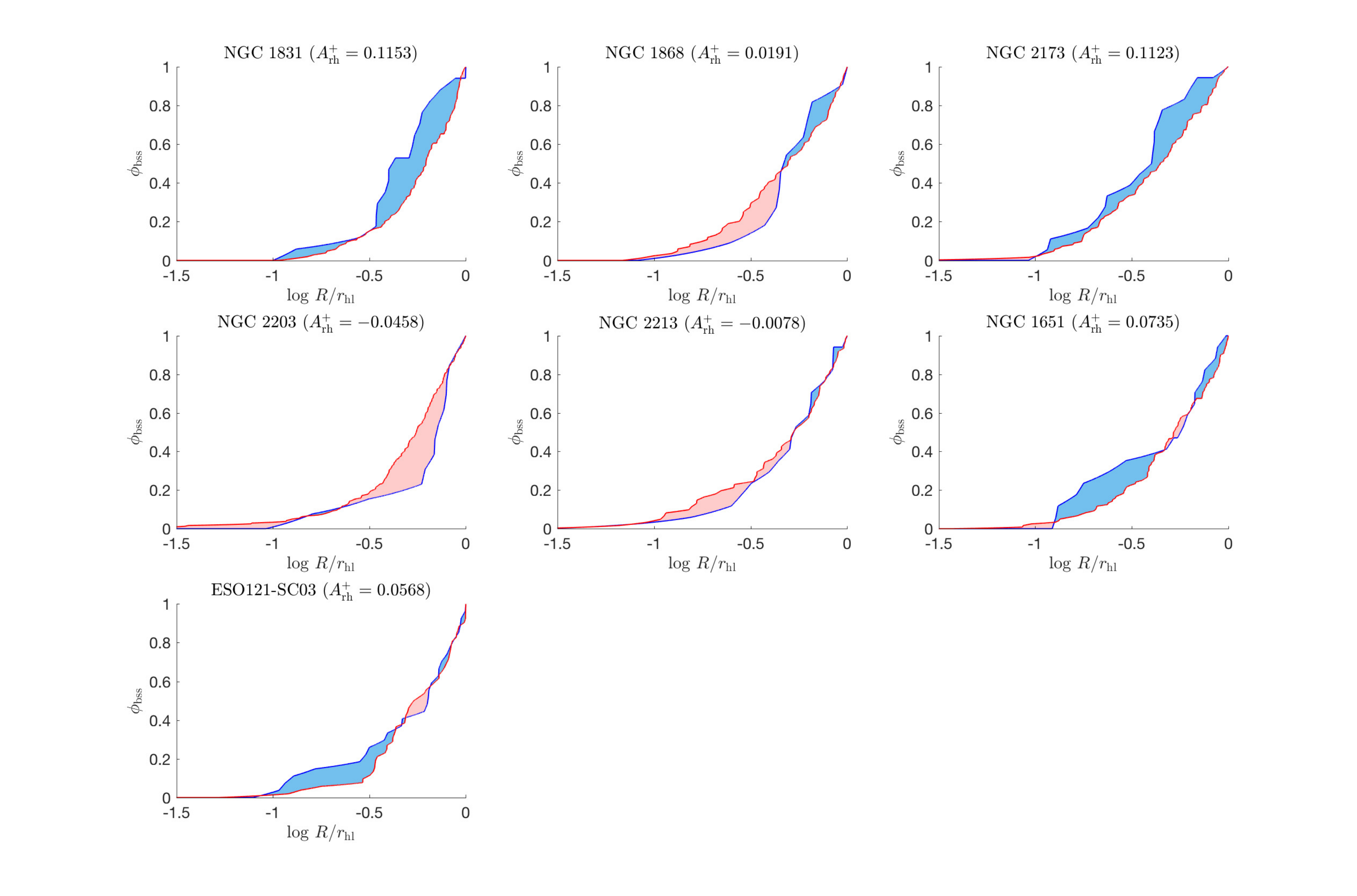}
\caption{Cumulative curves normalized to $r_{\rm hl}$ for BSSs (blue
  lines) and reference population stars (red lines). The area enclosed
  between these two curves is defined as $A^{+}_{\rm rh}$, as
  indicated by the blue shaded region (positive contribution to
  $A^{+}_{\rm rh}$) and pink shaded region (negative contribution to
  $A^{+}_{\rm rh}$).}
\label{F12}
\end{figure*}

For both the populations of BSSs and reference stars, we randomly
assigned $N_{\rm f}$ field stars to the cluster region for each
cluster. Here, $N_{\rm f}$ is drawn randomly from a normal
distribution centered on the expected number of field stars. The
spatial distribution of these artificial field stars is
homogeneous. We next subtracted these field stars from the
  observed sample of both BSSs and reference stars and calculated the
  `field-subtracted' cumulative profiles. We emphasize again that the
resulting radial distribution could strongly depend on the positions
of the subtracted field stars when the number of field stars is
small. Therefore, we repeated this procedure 1,000 times and
  adopted the average cumulative profiles of each sample as the final
  result. We then examined how this would change the resulting
$A^{+}_{\rm rh}$. Our final $A^{+}_{\rm rh}$ values are the
  averages of these 1,000 realizations. The associated uncertainty
  covers 95\% of all runs. Our results are illustrated in
  Fig. \ref{F12} and listed in Table \ref{T6}.

From Fig. \ref{F12} we find that the radial distributions of BSSs in
these LMC clusters vary. In NGC 1831, NGC 2173, NGC 1651, and
ESO 121-SC03, the BSSs are marginally more centrally concentrated than
the normal evolved stars. In NGC 1868, NGC 2203, and NGC 2213, the
radial distributions of the BSSs are very dispersed, showing no
evidence of apparent mass segregation. We estimated the dynamical ages
of the turnoff stars within the half-mass radii of our clusters. We
confirmed that they are all dynamically older than at least one
half-mass relaxation timescale (see the below). This is in contrast to
the GCs in the Milky Way, where \cite{Ferr18a} found that among 27
Galactic GCs observed as part of the {\sl HST} UV Legacy Survey, 17
have BSS $A^{+}_{\rm rh}$ greater than 0.18. We also applied the
  nonparametric $k$-sample ($k=2$) Anderson--Darling test to check
  whether the cumulative distributions of BSSs and reference
  population stars are drawn from the same underlying distribution
  (our null hypothesis). We found that we cannot reject the null
  hypothesis at a significance level of $p=0.05$.

\cite{Lanz16a} and \cite{Ferr18a} showed that $A^{+}_{\rm rh}$ can
measure the dynamical states of GCs in the Milky Way. In this paper,
we examined if their conclusion also hold for our sample of younger
GCs in the LMC. To do so, we uses the formula introduced by
\cite{Meyl87a} to calculate the half-mass relaxation timescales,
\begin{equation}
t_{\rm rh}=8.92\times10^5\frac{M_{\rm
    total}^{1/2}}{\bar{m}}\frac{r^{3/2}}{\log{(0.4M_{\rm
      total}/\bar{m})}}\,{\rm yr},
\end{equation}
where $M_{\rm total}$ is the cluster's total mass, $\bar{m}$ is the
typical mass for the stars of interest (in units of $M_{\odot}$), and
$r$ is the half-mass radius of interest. We assume that the half-light
radius derived here, $r_{\rm hl}$, is equal to the half-mass
radius. In principle, we have adopted a three-dimensional (3D) radius
under the assumption that our clusters are simple spherical systems.
In principle, the real 3D radius is 4/3 times the observed 2D
projected radius \citep{Spit87a}. $M_{\rm total}$ and $\bar{m}$ were
evaluated as follows. We first selected a sample of stars within the
half-light radius. For this stellar sample, we calculates the number
of stars within a range of F814W magnitude that is of order 2--3 mag
above the detection limit. We converted this magnitude range into a
stellar mass range by interpolation of the best-fitting isochrone. We
assumed that stars in this mass range follow a Kroupa mass function
\citep{Krou01a} and then evaluated the total number of stars by
extrapolating this mass function down to $0.08 M_{\odot}$. The total
mass for these stars thus represents half of the cluster total, since
we have assumed that the half-light radius is equal to the half-mass
radius. The average stellar mass is the total mass of all stars
divided by their total number.

Using the equation
\begin{equation}
t_{\rm r0}=1.55\times10^7\frac{1/\bar{m}}{\log{(0.5M_{\rm
      total}/\bar{m})}}v_0r_{\rm c}^2\, {\rm yr},
\end{equation}
we calculated the central relaxation time \citep{Meyl87a}. Here $v_0$
is the central velocity scale in units of km s$^{-1}$, which we
assumed to be equal to the core velocity dispersion,
\begin{equation}
v_0=\sigma_0=\sqrt{\frac{2GM_{\rm c}}{r_{\rm c}}}.
\end{equation}
The estimation of the core mass, $M_{\rm c}$, is similar to that of
the total mass.

In Table \ref{T6}, we present the calculated cluster total masses,
stellar average masses, half-light and central relaxation times, as
well as the BSSs' $A^{+}_{\rm rh}$. The associated uncertainties in
the clusters' total masses and their relaxation times are linked to
the uncertainties in the structural parameters. The average stellar
mass does not have an associated uncertainty, because it only depends
on the adopted mass function (we assumed a Kroupa mass function).

\begin{table*}
  \begin{center}
\caption{Calculated total cluster masses (2), average stellar masses
  (3), central relaxation times (4), half-light relaxation times (5),
  and $A^{+}_{\rm rh}$ for the BSSs (6).}\label{T6}
  \begin{tabular}{c c c c c c}\hline
    Cluster      &  $\log{M_{\rm tot}}$ & $\bar{m}$ & $\log{t_{\rm r0}}$ & $\log{t_{\rm rh}}$ & $A^{+}_{\rm rh}$ \\
   (1) & (2) [$M_{\odot}$] & (3)  [$M_{\odot}$] & (4) [yr] & [yr] (5) & (6) \\\hline
   NGC 1831 & $4.57^{+0.15}_{-0.12}$ & 0.37 & $8.86^{+0.14}_{-0.17}$ & $9.38^{+0.31}_{-0.19}$ & $0.1153^{+0.0269}_{-0.0336}$ \\
   NGC 1868 & $4.31^{+0.05}_{-0.11}$ & 0.35 & $8.21^{+0.10}_{-0.12}$ & $8.91^{+0.11}_{-0.19}$ & $0.0191^{+0.0312}_{-0.0109}$ \\ 
   NGC 2173 & $4.45^{+0.08}_{-0.15}$ & 0.34 & $8.79^{+0.11}_{-0.14}$ & $9.54^{+0.19}_{-0.31}$ & $0.1123^{+0.0341}_{-0.0321}$ \\ 
   NGC 2203 & $4.57^{+0.03}_{-0.05}$ & 0.33 & $9.28^{+0.05}_{-0.06}$ & $9.67^{+0.06}_{-0.10}$ & $-0.0458^{+0.0357}_{-0.0453}$ \\ 
   NGC 2213 & $4.57^{+0.04}_{-0.05}$ & 0.33 & $8.25\pm0.01$ & $9.06^{+0.09}_{-0.14}$ & $-0.0078^{+0.0491}_{-0.0492}$ \\ 
   NGC 1651 & $4.43\pm0.05$ & 0.33 & $9.06^{+0.08}_{-0.10}$ & $9.40^{+0.08}_{-0.10}$ & $0.0735^{+0.0370}_{-0.0369}$ \\ 
   ESO 121-SC03 & $3.92^{+0.08}_{-0.09}$ & 0.28 & $9.10^{+0.18}_{-0.23}$ & $9.33^{+0.12}_{-0.14}$ & $0.0568^{+0.0104}_{-0.0081}$ \\ 
    \hline
  \end{tabular} 
  \end{center} 
\end{table*} 

\cite{Lanz16a} measured the dynamical states of Galactic GCs based on
the logarithm of the ratio of a cluster's central relaxation
time and the Hubble time (13.7 Gyr), $\log{t_{\rm r0}/t_{\rm
    H}}$. However, this definition can only apply to GCs because
almost all Galactic GCs have ages close to the Hubble time. The
typical ages of our clusters span the range from less than 1 Gyr (NGC
1831) to $\sim$7 Gyr (ESO 121-SC03). Therefore, to appropriately
define the dynamical states of these younger GCs, we replaced the
Hubble time with their isochronal ages, $\log{t_{\rm r}/t_{\rm iso}}$,
where the isochronal ages are the ages associated with the
best-fitting isochrones for the bulk population stars.

\begin{figure}
\includegraphics[width=\columnwidth]{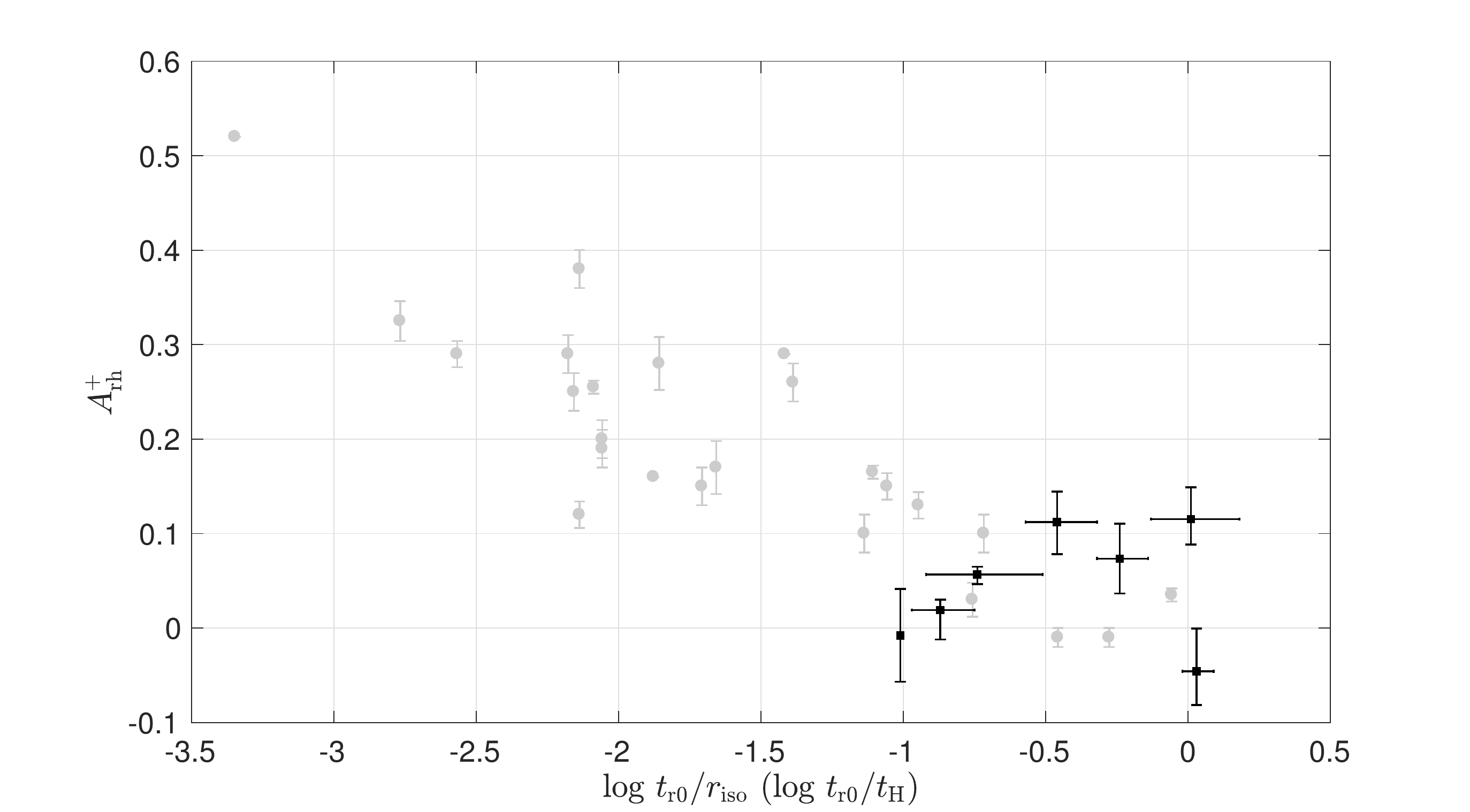}
\caption{$A^{+}_{\rm rh}$--$\log{t_{\rm r0}/t_{\rm iso}}$ for BSSs.
  The $A^{+}_{\rm rh}$--$\log{t_{\rm r0}/t_{\rm H}}$ diagram for the
  25 Galactic GCs \citep{Lanz16a} is also shown.}
\label{F13}
\end{figure}

In Fig. \ref{F13} we show the $A^{+}_{\rm rh}$--$\log{t_{\rm
    r0}/t_{\rm iso}}$ diagrams for BSSs in these seven clusters. To
make a direct comparison, the same diagram for 25 Galactic GCs is
plotted as well (note that for these clusters their $x$ positions are
$\log{t_{\rm r0}/t_{\rm H}}$). As shown in Fig. \ref{F13}, there is no
obvious correlation between $A^{+}_{\rm rh}$ and $\log{t_{\rm
    r0}/t_{\rm iso}}$ for these seven clusters. We have calculated the
Pearson coefficient between $A^{+}_{\rm rh}$ and $\log{t_{\rm
    r0}/t_{\rm iso}}$, which is 0.23. To check the relevant
significance, we generated seven points randomly distributed in the
diagrams and calculated their Pearson coefficient. We repeated this
procedure 10,000 times and counted how many times we obtained a
Pearson coefficient with a smaller absolute value than obtained for
the observations. We defined this count divided by 10,000 as the
significance of the correlations. This yields only 40\%. For
comparison, the Pearson coefficient for the correlation $A^{+}_{\rm
  rh}$--$\log{t_{\rm r0}/t_{\rm H}}$ for the 25 Galactic GCs studied
by \cite{Lanz16a} is $-0.85$, with a significance of 99\%. Clearly,
the correlation $A^{+}_{\rm rh}$--$\log{t_{\rm r0}/t_{\rm iso}}$ for
Galactic GCs is much tighter than for the LMC clusters. The lack of a
tight correlation between $A^{+}_{\rm rh}$ and $\log{t_{\rm r0}/t_{\rm
    H}}$ for these seven LMC clusters may simply be owing to their
dynamically young ages or to small-number statistics.

As shown in Fig. \ref{F13}, the distributions of $A^{+}_{\rm
  rh}$--$\log{t_{\rm r0}/t_{\rm iso}}$ for our LMC clusters and the
$A^{+}_{\rm rh}$--$\log{t_{\rm r0}/t_{\rm H}}$ for the Galactic GCs
overlap. All seven LMC clusters occupy the dynamically younger part of
the sequence defined by the Galactic GCs in the $A^{+}_{\rm
  rh}$--$\log{t_{\rm r0}/t_{\rm H}}$ diagram. If we combine our
results for BSSs in the LMC clusters with that for the Galactic GCs,
the resulting Pearson coefficient would still be $-0.84$.

\section{Physical Implications}\label{S4}

Since most of our clusters have smaller $A^{+}_{\rm rh}$ compared with
the Galactic GCs, they are also dynamically younger than the GCs. This
is consistent with \cite{Lanz16a} and \cite{Ferr18a}. Our results thus
imply that $A^{+}_{\rm rh}$ may be a potential `dynamical probe' for
extragalactic clusters at younger ages as well. To underpin this
conclusion, studying some globulars with extremely old dynamical ages
is essential. As an example, NGC 2019 is likely a core-collapse
cluster in the LMC \citep{Meyl87b}, which should be sufficiently
advanced dynamically (see Fig. \ref{F15}). Its BSSs should show a very
large $A_{\rm rh}^{+}$ if the empirical correlation as derived for
Galactic GCs also holds for the LMC clusters.

The $A^{+}_{\rm rh}$ for our clusters are more dispersed compared with
the values for Galactic GCs with equivalent dynamical ages, which may
be caused by their large uncertainties. For example, field
contamination is estimated by means of Monte-Carlo simulations rather
than direct observations of stellar proper motions. The latter method
has been employed for GCs since they are closer than the LMC clusters
\citep[e.g.,][]{Ferr18a}. The selection of our stellar samples may
also have an effect on the radial distributions of stars in some
clusters. In addition, the adopted structural parameters may affect
our results. We found that if we adopted the structural parameters
from the number density profiles rather than from the brightness
profiles, $A^{+}_{\rm rh}$ will change as well (although the
  change is very small indeed).

\begin{figure}
\includegraphics[width=\columnwidth]{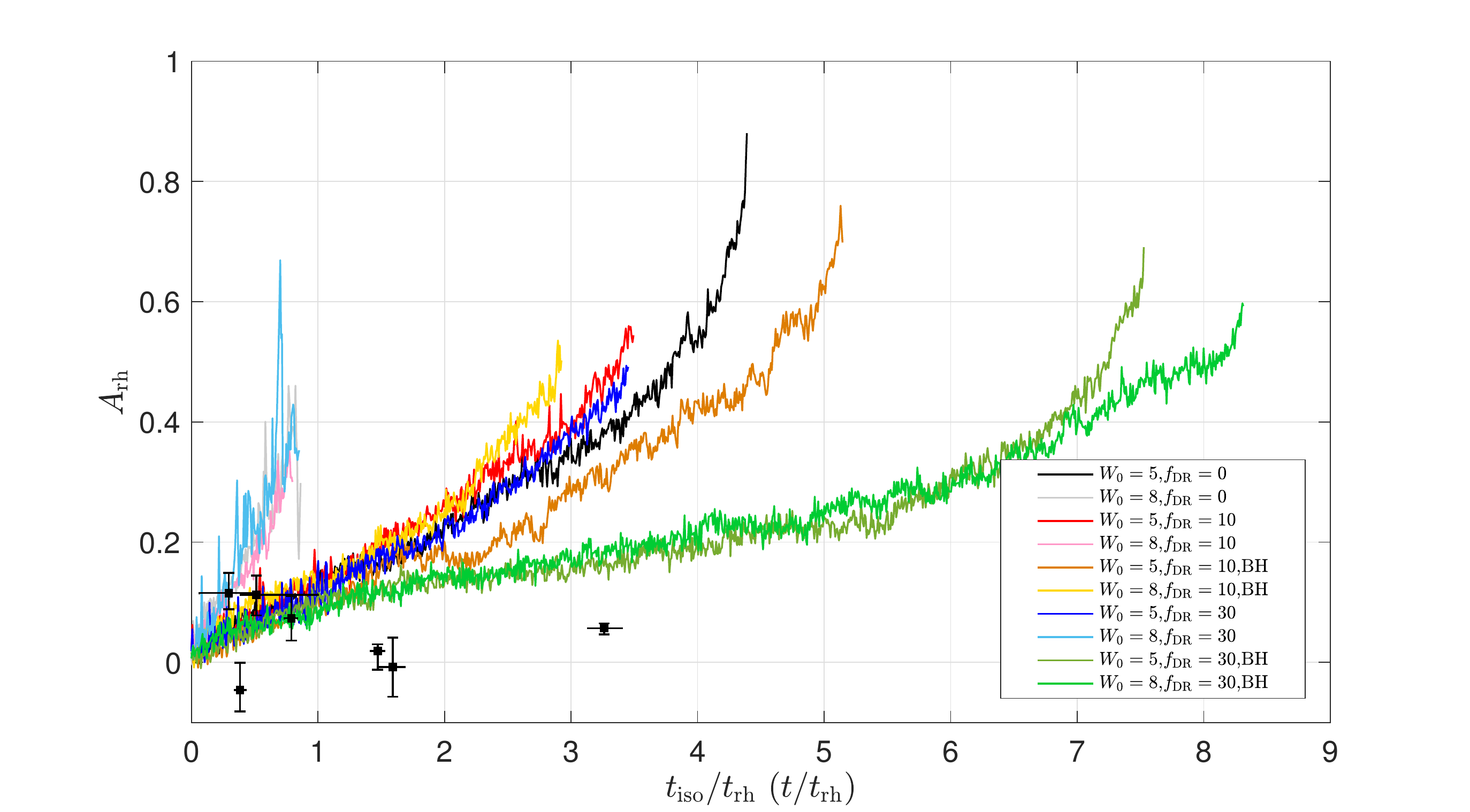}
\caption{$A^{+}_{\rm rh}$--$t_{\rm iso}/t_{\rm rh}$ diagram for BSSs
  (black squares). Evolutionary sequences of $A^{+}_{\rm rh}$ as a
  function of $t/t_{\rm rh}$ \citep{Ales16a} are shown as solid
  lines.}
\label{F14}
\end{figure}

\begin{figure}
\includegraphics[width=\columnwidth]{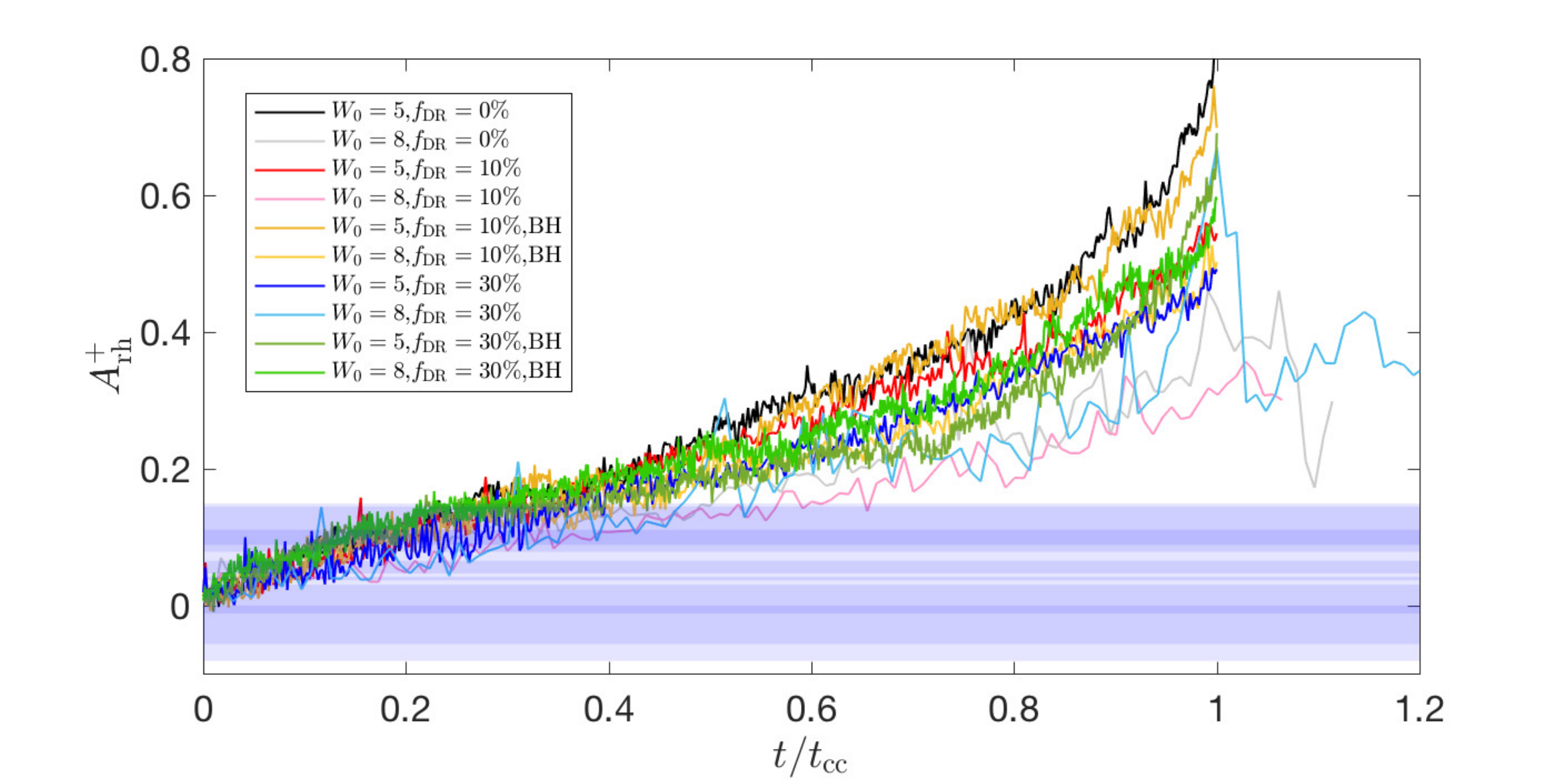}
\caption{As Fig. \ref{F15}, but with time normalized to the clusters'
  core-collapse time. The calculated $A^{+}_{\rm rh}$ ranges for the
  BSSs are shown as shaded regions.}
\label{F15}
\end{figure}

In Fig. \ref{F14}, we present a direct comparison with the numerical
simulations of \cite{Ales16a}. In this simulation, $A^{+}_{\rm rh}$
for the BSSs of star clusters is controlled by two physical
properties, the King central dimensionless potential, $W_0$, which
defines the initial central concentration and internal stellar
kinematics of star clusters, and the initial retention fraction of
dark remnants like neutron stars (NSs) or black holes (BHs), $f_{\rm
  DR}$. They found that, for a given $W_0$, the initial presence of a
large fraction of BHs will have a strong impact on the evolution of
$A^{+}_{\rm rh}$. As a result, stellar systems initially without BHs
will evolve more rapidly dynamically than those with BHs. The initial
concentration also affects $A^{+}_{\rm rh}$ at a given dynamical
age. $A^{+}_{\rm rh}$ is larger in stellar systems with a larger
initial concentration, although \cite{Ales16a} only ran two different
initial concentration models, characterized by $W_0=5$ and $W_0=8$.

As shown in Fig. \ref{F14}, only NGC 1831 may have a relatively high
degree of initial concentration level ($W_0=8$) and a negligible
initial retention fraction of BHs. Given (i) that the deviation
  of NGC 1831 is small and (ii) the positive result of the
  Anderson--Darling test, we conclude that the NGC 1831 data point may
  simply be caused by spurious noise. The other six clusters seem to
be consistent with models with a lower initial concentration ($W_0=5$)
and/or with an initial fraction of BHs ($f_{\rm DR}=$10, 30,
BH). Three of the clusters might be dynamically too young to show any
dynamical effects (younger than one half-mass relaxation time). As
shown by the simulation, except for the $W_0=8$, $f_{\rm DR}$=0;
$W_0=8$, $f_{\rm DR}$=10 and $W_0=8$, $f_{\rm DR}$=30 models, for the
other models there is almost no change in $A^{+}_{\rm rh}$ before one
half-mass relaxation timescale. For three other clusters which are
dynamically more advanced ($t_{\rm iso}/t_{\rm rc}>$1), $A^{+}_{\rm
  rh}$ is even significantly smaller than the values indicated by the
models.

In clusters with a small $A^{+}_{\rm rh}$, the radial distributions of
BSSs may be further affected by dynamical disruption of binary
stars. \cite{Gell13a} showed that due to the disruption of wide
binaries, the binary frequency will decrease toward the core of the
cluster after one crossing time. During the period from one crossing
time to one half-mass relaxation time, the radial distribution of
binaries will fall toward the cluster's core region. Subsequently,
their radial distribution will exhibit a bimodal morphology, similar
to that of BSSs in most GCs. If BSSs inherit the same dynamical
history as the binaries, it is possible that they will have a negative
$A^{+}_{\rm rh}$. If BSSs form later than most individual stars, their
radial distributions may have been shaped when they were born,
exhibiting a positive $A^{+}_{\rm rh}$. This is different from
\cite{Ales16a}. In their simulation $A^{+}_{\rm rh}$ is always zero or
positive. The binary dynamical disruption effect, combined with the
delayed mass segregation due to the presence of BHs, may lead to a
very small $A^{+}_{\rm rh}$ for a long period, which may explain why
some clusters have $A^{+}_{\rm rh}$ values that are even smaller than
those given by the models. Their low $A^{+}_{\rm rh}$ may also
indicate that the initial retention fraction of BHs is greater than
30\%. The combined effects of binary disruption and the presence of
BHs has been studied for NGC 2213 by \cite{Li18a}, who successfully
reproduced the non-segregated BSS population in this cluster. Here we
find $A^{+}_{\rm rh} = -0.0078$ for its BSSs, which means they are not
evolved dynamically at all. A comprehensive study of the effects of
BHs and binary disruption in other clusters will be explored in a
future study (J. Hong et al., in preparation).

We remind the reader that \cite{Ales16a} used the initial relaxation
time to scale the dynamical state of the clusters, while in this paper
we used the current relaxation time because the initial relaxation
time is not measurable. Some uncertainties may have been introduced
because of this. In general, we expect that a cluster will have a
shorter initial relaxation time because of the subsequent expansion
caused by evolutionary mass loss. If so, the actual dynamical age of
our clusters, if measured by their initial relaxation time, may be
larger than suggested by our current results.

\cite{Ales16a} showed that $A^{+}_{\rm rh}$ can be used to evaluate if
a cluster has experienced a core-collapse event. By comparing our
results with their simulation, we conclude that there is no
post-core-collapse cluster among our clusters. In Fig. \ref{F15} we
show the evolution of $A^{+}_{\rm rh}$ to the clusters' core-collapse
time \citep[see][their Fig. 6]{Ales16a}. The calculated ranges of
$A^{+}_{\rm rh}$ for the BSSs in our clusters are shown as the shaded
bands. Fig. \ref{F15} shows that when a star cluster reaches its
core-collapse phase, its $A^{+}_{\rm rh}$ will increase to at least
0.3, which is greater than $A^{+}_{\rm rh}$ derived for any of our
clusters. As shown in Fig. \ref{F15}, all clusters should be younger
than 40\% of the relevant core-collapse time. In summary, none of
these clusters are core-collapse clusters. As shown in \cite{Sun18a},
these clusters are not dense enough either to have high stellar
collision rates \citep{Chat13a}. Therefore, most of their BSSs may
have been formed through binary evolution.

\section{Summary and Conclusions}\label{S5}

We studied the radial distributions of BSSs in seven LMC clusters with
ages spanning from less than 1 Gyr to $\sim$7 Gyr. Using the same
method employed as \cite{Lanz16a} and \cite{Ferr18a}, we identified
their $A^{+}_{\rm rh}$ and determined their dynamical ages. We found
that our clusters are dynamically younger than the Galactic GCs
studied by \cite{Lanz16a} and \cite{Ferr18a}. Their $A^{+}_{\rm rh}$
are also smaller. The $A^{+}_{\rm rh}$--$\log{t_{\rm r0}/t_{\rm iso}}$
distribution of our clusters is consistent with that of the Galactic
GCs, indicating that the radial BSS distributions could potentially be
good indicators to measure the dynamical state of younger clusters.

By comparison of our results with the simulations run by
\cite{Ales16a}, we conclude that most of our clusters may initially
have contained a significant fraction of BHs. The presence of BHs may
have delayed their mass segregation, leading to a smaller $A^{+}_{\rm
  rh}$ than for star clusters initially without BHs. Our results show
that for many of our clusters $A^{+}_{\rm rh}$ is close to zero or
even negative, which means that their BSSs are almost unevolved
dynamically. We suggest that in addition to the presence of BHs,
dynamical binary disruption may have shaped the radial distributions
of the BSSs. The small $A^{+}_{\rm rh}$ values for our clusters also
indicate that none have experienced post-core-collapse events, which
is expected since our clusters are too young to go through core
collapse.

\acknowledgements

We thank Barbara Lanzoni at Bologna University for providing us with
the numerical simulation data. We thank Orsola de Marco at Macquarie
University for useful discussions. C. L. is supported by the Macquarie
Research Fellowship Scheme. This work was supported by the National
Key Research and Development Program of China through grant
2017YFA0402702 (RdG). J. H. acknowledges support from the China
Postdoctoral Science Foundation (grant 2017M610694). A. S. is
supported by the Natural Sciences and Engineering Research Council of
Canada. This work was partly supported by the National Natural Science
Foundation of China through grants U1631102, 11373010, 11473037,
11633005, and 11673032. Computations have been done in part on
computers of the Silk Road Project and the Laohu (Tiger) supercomputer
at the National Astronomical Observatories of the Chinese Academy of
Sciences (NAOC).

\facilities{{\sl Hubble Space Telescope} (WFC3/UVIS and ACS/WFC)}


\software{
{\sc dolphot2.0} \citep{Dolp16a} 
          }

\end{document}